\documentclass[11pt]{article}
\usepackage{graphicx}
\usepackage{epsfig}
\usepackage{latexsym,amsfonts,amsbsy,amssymb}
\usepackage{amsmath,amsthm}
\usepackage{color}
\usepackage{float} 
\usepackage[round, sort&compress, authoryear]{natbib}
\newtheorem{theorem}{Theorem}
\newtheorem{lemma}{Lemma}
\newcommand{\calL}{{\cal L}}
\newcommand{\E}{\mathbb E}
\newcommand{\prob}{\mathbb P}
\newcommand{\R}{\mathbb R}

\begin{document}
\title{HMC:  reducing the number of rejections by not using leapfrog and some results on the  acceptance rate}
\author{
M.P. Calvo\footnote{Departamento de Matem\'atica Aplicada e IMUVA, Facultad de Ciencias, Universidad de
Valladolid,  Spain. E-mail: mariapaz.calvo@uva.es} \,, D. Sanz-Alonso\footnote{Department of Statistics,
University of Chicago, 5747 South Ellis Avenue,  Chicago, Illinois 60637, US. E-mail: sanzalonso@uchicago.edu} \,
and J.M. Sanz-Serna\footnote{Departamento de Matem\'aticas, Universidad Carlos III de Madrid, Avenida de la
Universidad 30, E-28911 Legan\'es (Madrid), Spain. E-mail: jmsanzserna@gmail.com} }
 \maketitle
\begin{abstract}
The leapfrog integrator is routinely used within the Hamiltonian Monte Carlo method and its variants. We give strong numerical evidence that alternative, easy to implement algorithms yield fewer rejections with a given computational effort. When the dimensionality of the target distribution is high, the number of accepted proposals may be multiplied by a factor of three or more. This increase in the number of accepted proposals is not achieved by impairing any positive features of the sampling. We also establish new non-asymptotic and asymptotic  results on the monotonic relationship between the expected acceptance rate and the expected energy error. These results further validate the derivation of one of the integrators we consider and are of independent interest.
\end{abstract}
{\small{
{\bf{Keywords:}} Hamiltonian Monte Carlo, numerical integrators, expected acceptance rate, expected energy error.}}
\section{Introduction}

HMC, originally suggested by \cite{duane1987hybrid}, is a popular Markov chain Monte Carlo (MCMC) algorithm
where proposals are obtained by numerically integrating a Hamiltonian system of differential equations
\citep{neal2011,sanz2014markov}. This integration gives HMC its main potential advantage: the possibility of
proposals that are far away from the current state of the chain and, at the same time, may be accepted with
high probability. At present the leapfrog/St\"{o}rmer/Verlet scheme  is almost universally used to integrate
the Hamiltonian dynamics and obtain a proposal that may be accepted or rejected. While rejections increase the
correlation in any MCMC algorithm, they are particularly unwelcome in the
 case of HMC, because of the  computational effort wasted in the numerical integration legs
used to generate the rejected proposals \citep{extra}.

We emphasize that, when comparing different integrators, it is essential to take into account the
computational effort. If the numerical integration of the Hamiltonian dynamics were exact, then all proposals
would be accepted. By implication, for any reasonable (i.e.\ convergent) numerical integrator, it is possible
to make the acceptance rate arbitrarily close to \(100\%\) by reducing the step-length \(\epsilon\) and
increasing the number \(L\) of time-steps used to simulate the dynamics over a given time-interval \(0\leq
\tau\leq\tau_{end}\). However such a reduction in \(\epsilon\) may not be advisable as it implies a
corresponding increase in computational cost. In fact, \emph{once an integrator has been chosen}, a very high
acceptance rate signals that  an inappropriately small  value of \(\epsilon\) is being used and the
computational cost of generating each proposal is too high; one would sample better at the same cost exploring
the state space further by increasing the  number \(N\) of steps in the Markov chain and solving the Hamilton
equations less accurately in each proposal. On the other hand, \emph{when choosing an integrator} one should
prefer (all other things being equal) the method that maximizes the number of accepted proposals that may be
obtained with a given computational cost.

Section~\ref{sec:HMC} is introductory and contains a description of HMC and of the properties of the leapfrog
algorithm. It also serves to recap a number of considerations that are essential to understand the remainder
of the paper.

Section~\ref{sec:integrators} describes a two-parameter family of integrators that contains the leapfrog
algorithm as a particular case. All members of the family may be implemented as easily as leapfrog and all use
the same accept/reject strategy. It is not possible to identify analytically the algorithm of this family that
is best for the HMC method;  in fact the acceptance rate depends on the so-called global errors of the
numerical integration and it is well known that in  any real numerical simulation global errors
 (as distinct from local errors) are  extremely difficult to pin down
analytically (\cite{acta} provide a short introduction to the basic concepts in numerical integration).

 Motivated by the previous discussion, our first original contribution is to conduct thorough and systematic numerical experiments to compare different integrators (Section~\ref{sec:numerical}).
We employ three test problems: a simple multivariate Gaussian distribution, a Log-Gaussian Cox model and an example from
molecular dynamics: the canonical distribution of  alkane molecules. The last distribution possesses many
modes (corresponding to alternative equilibrium configurations of the molecule) and has been chosen in view of
the complexity of the nonlinear forces. In all test problems the dimensionality \(d\) of the target
distribution is high, ranging from 15 to 4096; in fact as \(d\) increases so does the computational cost of
generating proposals and therefore it is more important to identify efficient integrators. The experiments
show that several integrators of the family improve on leapfrog.  In particular  an integrator specifically designed
by \cite{blanes} to minimize within this family the expected energy error for univariate Gaussian targets, provides, for a given computational effort, three times more accepted proposals than leapfrog.

Section 5 contains theoretical results that complement our numerical findings and are also of independent interest. Our first main result (Theorem~\ref{th:acceptanceunivariate}) shows that, for univariate Gaussians, the expected acceptance rate is a monotonic function of the expected energy error. This implies that the integrator derived in \cite{blanes} to minimize the expected energy error indeed maximizes the expected acceptance rate. In  higher dimensional settings, we found in all our numerical experiments that the expected acceptance rate is approximately given by the formula \(a=2\Phi(-\sqrt{\mu/2})\),
where \(\mu\) is the mean energy error and \(\Phi\) the standard
normal cumulative distribution function. This formula appeared first in \cite{gupta1990acceptance} and its asymptotic validity was rigorously established by  \cite{optimal} under the very restrictive assumption that the target is a product of
independent identical copies of a given distribution. Our second main result (Theorem~\ref{th:central}) relaxes the restrictive product setting in \cite{optimal}, and in addition allows to vary along the high dimensional limit the integrator, the time-step and the length of the integration interval. These substantial extensions are possible by restricting the proof to multivariate Gaussian targets and building on the univariate Gaussian scenario considered in connection with Theorem~\ref{th:acceptanceunivariate}. In summary, the theoretical contributions in Section 5 provide new strong evidence that the only way to improve the average acceptance rate is by designing integrators that reduce the expected energy errors.

We conclude this introduction with some caveats. A comparison of the merits of HMC with those of other
sampling techniques or of the advantages of the various variants of HMC is completely outside our scope here;
a recent contribution devoted to such comparisons is the paper by  \cite{radivojevic2019modified}, which also
provides an extensive bibliography. Similarly, we are not concerned with  discussing issues of HMC not
directly related to inaccurate time-integrations. Some of those issues concern the ergodicity of the method
\citep{cances,livingstone2019geometric,randomized},  the choice of the length \(\tau_{end}\) of the
time-interval for simulating the dynamics \citep{hoffman2014no}, the difficulties in leaving the neighbourhood
of a probability mode in multimodal targets \citep{mangoubi2018does} and the scalability of the method to
large data-sets \citep{chen2014stochastic}. Finally, while we limit the exposition to the simplest version of
HMC, our study may be applied to other variants
\citep{horowitz1991generalized,akhmatskaya2017adaptive,radivojevic2018multi} and even to molecular dynamics
simulations with no accept/reject mechanism in place \citep{fernandez2016adaptive}.  Among the variants of HMC, NUTS
\citep{hoffman2014no} is the most widely used by statisticians, as it frees the user from the burden of tuning the time-integration parameters. Referring to the formulation in the original paper \citep{hoffman2014no} (but there are alternative formulations),
NUTS includes a kind of delayed-rejection procedure where the value of the target density after each single time-step is tested against an appropriate random variable \(u\). Energy errors reduce the number of points  that pass that test and therefore the number of locations the Markov chain may jump to at each step. In addition one would expect that points farther away from the initial condition are more likely to suffer from high energy errors and therefore more accurate integrations would imply larger jumps in the Markov chain.
The integrators considered here may certainly be incorporated to NUTS; a careful evaluation of the advantages of such an incorporation may be the subject of future research.

\section{Hamiltonian Monte Carlo}
\label{sec:HMC}

We denote by  \(\theta\in\R^d\) the random variable we wish to sample from,  \(\pi(\theta)\) the corresponding density  and
 \(p\in\R^d\) the auxiliary variable with density \({\cal N}(p|0,M)\).
 The joint density and the negative log-likelihood are
then, respectively, \(\pi(\theta,p)=\pi(\theta){\cal N}(p|0,M)\) and
\[
H(\theta,p) = -\calL(\theta)+\frac{1}{2}p^TM^{-1}p+\frac{1}{2} \log\big( (2\pi)^d \det(M)\big),
\]
where \(\calL(\theta)= \log \pi(\theta)\).\footnote{The constant \((1/2) \log\big( (2\pi)^d \det(M)\big)\) may
be suppressed without altering the Hamiltonian dynamics or the acceptance probability. } Hamilton's differential equations associated with  the energy \(H\)  are  given by
\begin{equation}\label{eq:hamdynamics}
\frac{d}{d\tau} \theta= +\frac{\partial H}{\partial p}=M^{-1}p,\qquad
\frac{d}{d\tau} p = -\frac{\partial H}{\partial \theta} = \nabla_\theta \cal L(\theta),
\end{equation}
where \(\tau\) represents time and \(M\) is the mass matrix.  HMC  integrates  the system \eqref{eq:hamdynamics} by means of a numerical integrator whose error is removed by an accept/reject mechanism. At present the
 leapfrog/St\"{o}rm\-er/Verlet algorithm is the integrator of choice. Over a single time-step
 of length \(\epsilon>0\), the integrator reads
\begin{eqnarray}
\label{eq:leapfrog1} p & \leftarrow & p+\frac{\epsilon}{2}  \nabla_\theta \cal L(\theta),\\
\label{eq:leapfrog2} \theta & \leftarrow & \theta+\epsilon M^{-1}p,\\
\label{eq:leapfrog3} p & \leftarrow & p+\frac{\epsilon}{2}  \nabla_\theta \cal L(\theta).
\end{eqnarray}

The bulk of the computational effort typically  lies in the evaluation
 of the gradient (score) \(\nabla_\theta\calL(\theta)\). While, in a single time-step, an evaluation
 is required by
\eqref{eq:leapfrog1} and a second  evaluation is required by \eqref{eq:leapfrog3}, an integration leg
consisting of \(L\) time-steps only demands \(L+1\) evaluations, because the gradient in  \eqref{eq:leapfrog3}
is the same as the gradient in \eqref{eq:leapfrog1} at the next time-step.  Thus the computational cost of the
leapfrog integrator is essentially one gradient evaluation per time-step.

At each step of the Markov chain  the Hamiltonian dynamics over an interval \(0\leq \tau \leq \tau_{end}=
L\epsilon\) are simulated by taking \(L\) time-steps of length \(\epsilon\) of the integrator starting from
\((\theta,p)\), where  \(\theta\) is the last sample and \(p\) is drawn from \({\cal N}(p|0,M)\). The numerical
solution \((\theta^\star,p^\star)\) at the end of the \(L\) time-steps is
  a (deterministic) proposal,  which is accepted with probability
\begin{equation}\label{eq:acceptance}
a=\min\Big(1,\exp\big(-[H(\theta^\star,p^\star)-H(\theta,p)]\big)\Big).
\end{equation}
It is hoped that, by choosing \(\tau_{end}\) appropriately, \(\theta^\star\) will be far from  \(\theta\) so as
to reduce correlation and enhance the exploration of the state space.

In  \eqref{eq:acceptance}, the difference
\begin{equation}\label{eq:difference}
   \Delta H(\theta,p) = H(\theta^\star,p^\star)-H(\theta,p)
\end{equation}
(recall that \(\theta^\star\) and \(p^\star\) are deterministic functions of \((\theta,p)\)) is the
\emph{increment} of the Hamiltonian function over the numerical integration leg. For the exact solution, the
value of \(H\) remains constant as \(\tau\) varies  and therefore the \emph{initial}
energy \(H(\theta,p)\) coincides with the  final energy at \(\tau_{end}\). Thus, \eqref{eq:difference} is also the \emph{error} in \(H\) at \(\tau_{end}\)
introduced by the numerical integrator. Since leapfrog is a \emph{second order method}, in the limit where
\(\epsilon \rightarrow 0\) and \(L\rightarrow \infty\) with \(L\epsilon=\tau_{end}\) fixed,  it provides
numerical approximations to the true values of the Hamiltonian solutions that have \(\mathcal{O}(\epsilon^2)\)
errors.\footnote{This and similar estimates to be found later require some smoothness in the Hamiltonian but
we will not concern ourselves with such issues.} It follows that, for fixed \((\theta,p)\),
\eqref{eq:difference} is also \(\mathcal{O}(\epsilon^2)\) in that limit. As a consequence, once \(\tau_{end}\)
has been chosen to generate suitable proposals, the acceptance probability \eqref{eq:acceptance} may become
arbitrarily close to \(100\%\) by decreasing \(\epsilon\) (but this will imply an increase of the
computational cost of generating the proposal, as more time-steps have to be taken to span the interval
\([0,\tau_{end}]\)). The sentence \lq\lq HMC may generate proposals away from the current state that may be
accepted with high probability\rq\rq\ has been repeated again and again to promote HMC.

For given \(\tau_{end}\), how should \(\epsilon\) be chosen? If \(\epsilon\) is too large, many rejections
will take place and the chain will be highly correlated. On the other hand, a very high empirical acceptance
rate signals that  an inappropriately small  value of \(\epsilon\) is being used and the computational cost of
generating each proposal is too high; one would sample better at the same cost by  increasing the  number
\(N\) of steps in the Markov chain and solving the Hamilton equations less accurately in each proposal.
According to the analysis in \cite{optimal}, that only applies in a restrictive model scenario where the
target is a product of independent copies of a fixed distribution, one should tune \(\epsilon\) so as to
observe an acceptance rate of approximately \(65\%\). The result is asymptotic as \(d\uparrow \infty\) and
that reference recommends that in practice one aims at acceptance rates larger than that. This discussion will
be retaken below, when commenting on Figure \ref{Gaussian_fig3_256}.

 In an HMC context numerical integrations do not need to be very accurate. Even
 if the energy error in an integration leg is as large as, say, \(+0.5\) (orders of magnitude larger
 than the accuracy usually sought when integrating numerically differential equations), then the proposal
  will be accepted with  probability \(\exp(-0.5)\approx 0.61\). In addition, negative energy errors always result in acceptance. Nevertheless it is important to recall that the Hamiltonian is an extensive quantity whose value is added when mechanical systems are juxtaposed. If the vector random variable \(\theta\) has \(d\) independent  scalar components and  \(M\) is diagonal, then the Hamiltonian for \(\theta\) is the sum of the Hamiltonians for the individual components \((\theta_i,p_i)\) and the Hamiltonian system of
   differential equations for \((\theta,p)\) is the uncoupled juxtaposition of the one-degree-of-freedom Hamiltonian equations for the \((\theta_i,p_i)\).
  In that model scenario, an acceptance probability of \(\exp(-0.5)\approx 0.61\) would require
  that each one-degree-of-freedom Hamiltonian system be integrated with an
  error of size \(0.5/d\) (on average over the components). Thus, as \(d\) increases the numerical integration has
  to become more accurate.

The leapfrog integrator has two  geometric properties \citep{sanz2018numerical,blanes2016concise}: (1) it
preserves the volume element and (2) it is time-reversible.
These two properties are essential for  the accept/reject mechanism based on the expression
\eqref{eq:acceptance} to remove the bias introduced by numerical integration (see \cite{compressible}) for integrators requiring more complicated accept/reject rules).

To conclude this section, we look at the model problem with univariate target \(\pi(\theta) = {\cal
N}(\theta|0,\sigma^2)\) and \(M=1\), so that \(\pi(p) =  {\cal
N}(p|0,1)\). For this problem, it is proved by \cite{blanes} that the expectation at stationarity of the difference
\eqref{eq:difference}, for \(0<\epsilon<2\sigma\) and an arbitrary number \(L\) of time-steps, satisfies
\begin{equation}\label{eq:expectationbound}
0\leq \E(\Delta H) \leq \frac{\epsilon^4/\sigma^4}{32\big(1-\frac{\epsilon^2}{4\sigma^2}\big)}.
\end{equation}
 With \(\epsilon = \sigma\), the upper bound of the energy error takes the value \(1/24\) and halving \(\epsilon\) brings down the expected error  to \(\leq 1/480\). This proves  that,
 for this model problem, the leapfrog integrator works well with step-lengths that are not small (as measured in the \lq\lq natural\rq\rq\ unit \(\sigma\)).

The bound \eqref{eq:expectationbound} requires that \(\epsilon/\sigma\in(0,2)\). In fact for
\(\epsilon/\sigma\geq 2\) the leapfrog algorithm is \emph{unstable} for the corresponding Hamiltonian
equations given by the (scalar) harmonic oscillator \( d\theta/d\tau = p\), \(dp/d\tau = -\theta/\sigma^2\). Recall that this means
that, if for fixed \(\epsilon\geq 2\sigma\) more and more time-steps are taken, then the numerical solution
grows unboundedly. One says that \((0,2)\) is the \emph{stability interval} of the  integrator.

As pointed out before,  as \(\epsilon\downarrow 0\)  with \(L\epsilon = \tau_{end}\), the energy increment
\eqref{eq:difference} is \(\mathcal{O}(\epsilon^2)\) at each fixed \((\theta,p)\); since, according to
\eqref{eq:expectationbound}, its expectation at stationarity is \(\mathcal{O}(\epsilon^4)\) , there is much
cancellation between points \((\theta,p)\) where  \eqref{eq:difference} is positive and points where it is
negative. This cancellation holds for arbitrary targets \citep{optimal,acta}.

\section{A family of integrators}
\label{sec:integrators}
In this paper we work with a family of splitting integrators \citep{blanes2008splitting}, depending on two
real parameters \(b\) and \(c\), that have  potential for replacing the standard leapfrog method within HMC
algorithms. For the analysis of this family, see \cite{campos}.
 The formulas for
performing a single time-step are:
\begin{eqnarray}
\label{eq:new1}   p &\leftarrow&p+(1/2-b)\epsilon\, \nabla_\theta \cal L(\theta),  \\
   \theta &\leftarrow&\theta + c\epsilon M^{-1}p, \\
   p &\leftarrow& p+b\epsilon\, \nabla_\theta \cal L(\theta), \\
   \theta &\leftarrow&\theta+ (1-2c)\epsilon M^{-1}p,  \\
   p &\leftarrow& p+b\epsilon\, \nabla_\theta \cal L(\theta), \\
   \theta &\leftarrow&\theta+ c\epsilon M^{-1}p,  \\
 \label{eq:new7}  p &\leftarrow& p+(1/2-b)\epsilon\, \nabla_\theta \cal L(\theta).
 \end{eqnarray}
  We  assume that \(b\) and \(c\) do not take the values \(0\) or \(1/2\); when they do some of the substeps are redundant.
 An integration based on this algorithm is just a sequence of kicks and drifts and therefore not very different from a standard leapfrog integration.
 Four evaluations of the gradient are required at a single time-step \eqref{eq:new1}--\eqref{eq:new7}. However, since the last gradient at a time-step may be reused at the next time-step, an integration leg of \(L\) time-steps needs \(3L+1\) gradient evaluations. Therefore, over a single time-step \eqref{eq:new1}--\eqref{eq:new7} is \emph{three} times more expensive than the leapfrog \eqref{eq:leapfrog1}--\eqref{eq:leapfrog3}. \emph{To have fair comparisons, an HMC simulation based on the standard leapfrog \eqref{eq:leapfrog1}--\eqref{eq:leapfrog3} should be allowed three times as many time-steps as a simulation based on \eqref{eq:new1}--\eqref{eq:new7}.} We will return to this issue later.

Since each individual substep in \eqref{eq:new1}--\eqref{eq:new7}  preserves volume, the integrator \eqref{eq:new1}--\eqref{eq:new7} is
volume-preserving. In addition, its palindromic structure leads to its being time-reversible. As a
consequence, the bias introduced by numerical integration errors may be suppressed by an accept/reject
mechanism based on the same formula \eqref{eq:acceptance} that is used for leapfrog integrations. In summary,
the implementation of HMC based on the new integrator is extremely similar to leapfrog-based implementations.

\cite{campos} show that when the parameters do not satisfy the relation
\begin{equation}\label{eq:constraint}
  b+c-6bc= 0,
\end{equation}
the  stability interval of \eqref{eq:new1}--\eqref{eq:new7} is relatively short and as a consequence small
values of \(\epsilon\) are required to cover the interval \([0,\tau_{end}]\). Therefore schemes for which
\eqref{eq:constraint} is violated cannot possibly compete with standard leapfrog. Accordingly, we hereafter
assume that the relation holds and in this way we are left with a family that may be described in terms of the
single parameter \(b\). It is perhaps of some interest to point out that by imposing \eqref{eq:constraint} we
exclude the unique choice of \(b\) and \(c\) for which \eqref{eq:new1}--\eqref{eq:new7} is fourth-order
accurate. That fourth-order accurate integrator has a very short stability interval and performs extremely
poorly in the HMC  context, as shown by \citet{campos}. All integrators considered below are then
\emph{second-order accurate}, just as standard leapfrog.

In the analysis of numerical integrators, it is well known that it is virtually impossible to get analytically useful information on global errors (except  for very simple models like the harmonic oscillator). In particular, in our context it is not possible to obtain useful information on the energy increment
\eqref{eq:difference} for a given Hamiltonian system, integrator and value of \(\epsilon\). This is even more so
when,  for reasons explained in the preceding section, we are interested in \emph{large} values of \(\epsilon\).
 Therefore
numerical experimentation is essential to identify good values of \(b\). That experimentation has necessarily to be limited to particular choices  of \(b\) and
  the numerical experiments reported in the next section compare  six values of \(b\) that are representative.
   The six choices of \(b\) fall into two categories. We include three choices (denoted by LF, BlCaSa and PrEtAl)
    that correspond to \lq\lq principled\rq\rq\ choices of \(b\), and we supplement them with three other
     values corresponding to \lq\lq round figures\rq\rq\ in the interval \([0.33, 0.45]\). Experiments not reported here clearly show that values of \(b\) smaller than \(0.33\) or larger than \(0.45\) lead to integrators that
 are not competitive in the HMC context with the six listed below. As we shall see in later sections, the relative performance of the six integrators turns out not to essentially depend on the distribution being sampled.\footnote{It is also independent of the choice of mass matrix, but all experiments reported use the unit mass matrix.}

\begin{itemize}
\item \(b = 1/3\). In this case, it is  easy to check that a single time-step of
    \eqref{eq:new1}--\eqref{eq:new7} yields the same result as three consecutive time-steps of standard
    leapfrog \eqref{eq:leapfrog1}--\eqref{eq:leapfrog3} with step-length \(\epsilon/3\). Thus, in the
    experiments below, one may think that, rather than performing \(L\) time-steps with
    \eqref{eq:new1}--\eqref{eq:new7}, each of length \(\epsilon\), one is taking \(3L\) time-steps with
    length \(\epsilon/3\) with the standard leapfrog integrator. Hence, when comparing later  an integration
    with a value \(b\neq 1/3\) with  a second integration with \(b=1/3\), we are really comparing the
    results of the first integration with those standard leapfrog would deliver if it were run with a
    time-step \(\epsilon/3\), so as to equalize computational costs. With \(b=1/3\) the length of the
    stability interval of \eqref{eq:new1}--\eqref{eq:new7} is \(\eta=6\) (of course three times the
    corresponding length for \eqref{eq:leapfrog1}--\eqref{eq:leapfrog3}). In what follows we shall refer to
    the method with \(b=1/3\) as LF.
\item \(b= 0.35\) with stability interval of length \(\eta\approx 4.969\).
\item \(b=0.38111989033452\). The rationale for this choice, suggested by \cite{blanes} in the numerical
    analysis literature, will be summarized below. This method is called BlCaSa in \cite{campos}. The length
    of the stability interval is \(\eta \approx 4.662\).
\item \(b = 0.391008574596575\). This choice, due to  \cite{predescu} in the molecular dynamics literature
    without reference to HMC, results from imposing that, for the particular case where \(\nabla
    \calL(\theta)\) is linear (so that \(\pi(\theta)\) is a  Gaussian), the error in \(H(\theta,p)\) after a
    single time-step is \(\mathcal{O}(\epsilon^5)\) as \(\epsilon\downarrow 0\) (rather than
    \(\mathcal{O}(\epsilon^3)\) as it is the case for other values of \(b\)). As a result, the energy
    increment \eqref{eq:difference} over an integration leg is \(\mathcal{O}(\epsilon^4)\) \emph{for
    Gaussian targets}, which, according to the results in \cite{optimal}, implies an
    \(\mathcal{O}(\epsilon^8)\) bound for the expected energy error for such targets. This method is called
    PrEtAl in \cite{campos} and has a stability interval of length \(\eta \approx 4.584\).
\item \(b = 0.40\), with \(\eta\approx 4.519\).
\item \(b = 0.45\), with \(\eta\approx4.224\).
\end{itemize}

As it is the case for PrEtAl, the method BlCaSa was derived assuming a  Gaussian model. However,
while PrEtAl was derived so as to reduce the energy error in the limit \(\epsilon\downarrow 0\), the
derivation of BlCaSa takes into account the behaviour of the method over a finite \(\epsilon\)-interval in
view of the fact that Hamiltonian simulations within HMC are not carried out with small values of the
step-length. It was proved by \cite{blanes} that for each integrator, assuming \(M=I\),\footnote{The result
may be adapted to cover other choices of mass matrix \citep{blanes,acta}.} the expected energy increment at
stationarity has, independently of the number \(L\) of time-steps at each integration leg, a bound
\begin{equation}\label{eq:bound}
0\leq \E\big(\Delta H\big) \leq \sum_{i=1}^d \rho(\epsilon /\sigma_i),
\end{equation}
where \(d\) is the dimension of \(\theta\), the \(\sigma_i\) are the marginal standard deviations (i.e.\ the
square roots of the eigenvalues of the covariance matrix) of \(\theta\) and \(\rho\) is a function that
changes with the integrator. For the standard leapfrog integrator and \(d=1\), we have seen this bound in
\eqref{eq:expectationbound}. \cite{blanes}  determined \(b\) by minimizing
\[
\rho_\infty = \max_{0<\zeta<3} \rho(\zeta);
\]
this optimizes the performance assuming that \(\epsilon\) is chosen \(\leq 3 \min_i \sigma_i\),
 see
\cite{blanes} for  further details.\footnote{For the standard leapfrog integrator the step-length \(\epsilon=
\min_i \sigma_i\) is \(50\%\) smaller than the maximum allowed by stability (compare with
\cite{neal1993,neal2012}); the interval in the minimization here is three times as long because the
integrators  should be operated
 with step-lengths three times longer than \eqref{eq:leapfrog1}--\eqref{eq:leapfrog3}.}
\section{Numerical experiments}
\label{sec:numerical}
Appropriate choices of \(M\) may allow for proposal mechanisms tailored to the given target and to its
possibly anisotropic covariance structure.
 From the point of view of the Hamiltonian dynamics, \(M\) may be chosen so as to reduce
 the spectrum of frequencies present in the  system of differential equations, which facilitates the numerical integration, see e.g.\ \cite{hilbert}. For simplicity we do not explore
  here special choices of \(M\) and
all numerical experiments use the unit mass matrix \(M=I_d\).
In all of them we have randomized, using the recipe in \cite{neal2011}, the step-length at the beginning of each step of the Markov chain/integration leg. Once the \lq\lq basic\rq\rq\ step-length \(\epsilon\)  and the number of time-steps \(L\) to complete a simulation of duration \(\tau_{end} = L\epsilon\) have been chosen,  the integration to find the \(n\)-th proposal of the chain is performed with a step-length
\[
\epsilon^{(n)} = (1+u^{(n)})\epsilon, \quad u^{(n)} \sim {\cal U}(-0.05, 0.05).
\]
As a result, the actual length  \(L\epsilon^{(n)}\) of the integration interval will differ slightly from \(\tau_{end}\).
 This randomization suppresses  the artifacts that may appear for   special choices
 of the step-length \citep{randomized}.

\subsection{A Gaussian example}
We have first considered a simple test problem, used by \cite{blanes}, with target density
\begin{equation}\label{eq:gaussianexample}
\pi(\theta)\propto\exp{\left ( -\frac{1}{2} \sum_{j=1}^d j^2 \theta^2_j \right )},
\end{equation}
for \(d=256\) or \(d=1024\). Assuming a diagonal covariance matrix is \emph{not restrictive}, as the application of the numerical integrators commutes with the rotation of axes that diagonalizes the covariance matrix (see \cite{acta} for a detailed discussion). We ran several values of \(\tau_{end}\); the performance of each integrator varies with \(\tau_{end}\) but the conclusions as to the merit of the various integrators do not. We report the case \(\tau_{end}=5\) because it is representative, in the sense that integrators do not show their best or worst performances.

As in this example the Hamiltonian system to be integrated is a set of \(d\) uncoupled linear oscillators with angular frequencies \(1,\dots,d\), the condition for a stable integration is
\[d \times \epsilon \leq \eta,\]
where \(\eta\) is the length of the stability interval of the method. For all six integrators tested, \(\eta\geq 4\) and therefore
for the step-length \(\epsilon_0 = 4/d\) all of them are stable.  For the case \(d= 256\),
and \(\tau_{end} = 5\), \(\epsilon_0 = 4/d\), we need \(L = \tau_{end}/\epsilon_0 = 320\) time-steps in each integration leg. Our experiments had \(L = 320, 360, 400, \dots, 960\) and \(\epsilon = 5/320, 5/360, \dots , 5/960\),
 respectively; (\(\epsilon\) then varies between \(\approx 1.6\times 10^{-2}\) and \(\approx 5.2\times 10^{-3}\)). For the methods with longer stability intervals, such as LF, we also used the coarser values \(\epsilon= 5/280, 5/240, 5/200\), with (respectively)
\(L = 280, 240, 200\). Turning now to \(d= 1024\), the choices  \(\tau_{end} = 5\), \(\epsilon_0 = 4/d\), require
\(L = \tau_{end}/\epsilon_0 = 1280\) time-steps in each integration leg. Our experiments used \(L = 1280, 1440, 1600, \dots,\) \( 3840\), \(\epsilon = 5/1280, \dots, 5/3840\) (\(\epsilon\) ranges from \(\approx 3.5\times 10^{-3}\) to
\(\approx 1.3\times 10^{-3}\)), and in addition \( L = 1120, 960, 800\), \(\epsilon = 5/1120, 5/960, 5/800\) for the integrators with longer stability intervals.

For each \(b\) and \(\epsilon\), we have generated a  Markov chain with \(N=5000\) samples, a number that is
large enough for our purposes. The initial \(\theta^{(0)}\) is drawn from the target \(\pi(\theta)\), and we
have monitored the acceptance rate, the mean, across the \(5000\)
integration legs/Markov chain steps, of the energy increments \eqref{eq:difference} and the following metrics:
\begin{itemize}
\item The effective sample size (ESS) of the first component \(\theta_1\)  of
\(\theta\). This is the component with largest standard deviation \(\sigma_1 =1\) and also has the smallest ESS values. (The behaviour of the ESS and average jump distance for this problem, in the absence of numerical integration error, is analyzed by \cite{randomized}.)
\item The ESS for the components \(\theta_{d/2}\) and \(\theta_d\) with standard deviations \(2/d\) and \(1/d\) respectively.
\item The ESS for the positive variables \(\theta_1^2\), \(\theta_{d/2}^2\) and \(\theta_d^2\). This
    complements the preceding metrics, since the ESS of the \(\theta_i\), with symmetric distributions, may
    increase due to negative autocorrelations.
\item The average square jump distance.
\end{itemize}

It is important to note that we are not comparing six essentially different sampling algorithms: different values of \(b\) merely lead to not very  different integration errors of one and the same Hamiltonian dynamics. Accordingly, for a given initial state, the proposals generated by the six integrators are  close to one another (and to the exact solution of the dynamics) and therefore the six algorithms should not differ much  in their  exploration of  the state space. However a very small difference
 in energy error between integrations with the same initial condition and two different values of \(b\) may imply that in one case the proposal is accepted and in the other rejected. For this reason  it is reasonable to expect that the behaviour of the metrics listed above is determined by the acceptance rate. The extensive numerical experiments we performed show that this is indeed the case.
 For brevity we shall only report results on the first of the metrics listed above; other metrics lead to the same conclusions both as to the behaviour of the metric as \(\epsilon\) decreases for a given \(b\) and as to the relative merit of the different values of \(b\).

\begin{figure}
\begin{center}
\includegraphics[width=0.48\hsize]{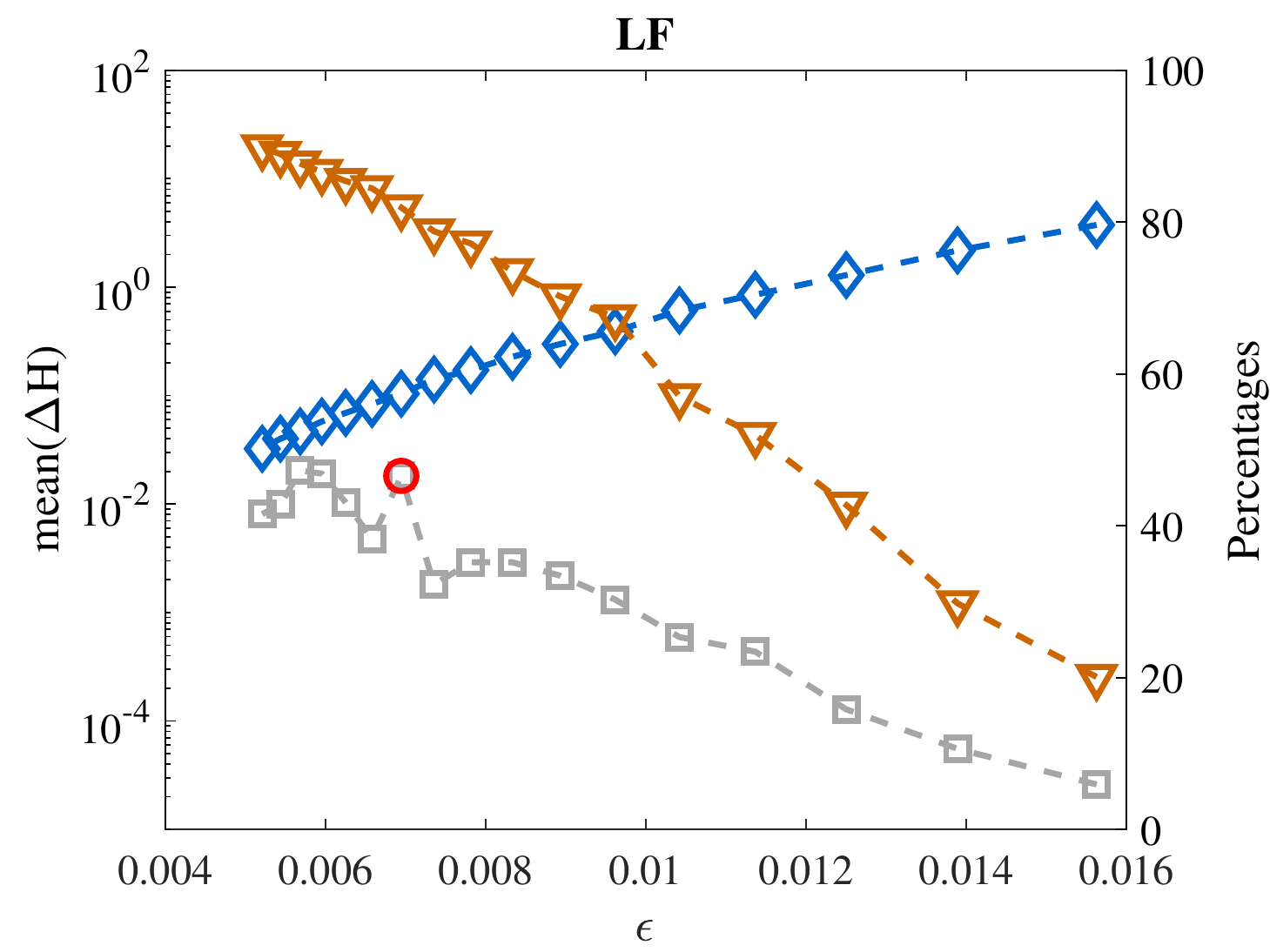}
\quad
\includegraphics[width=0.48\hsize]{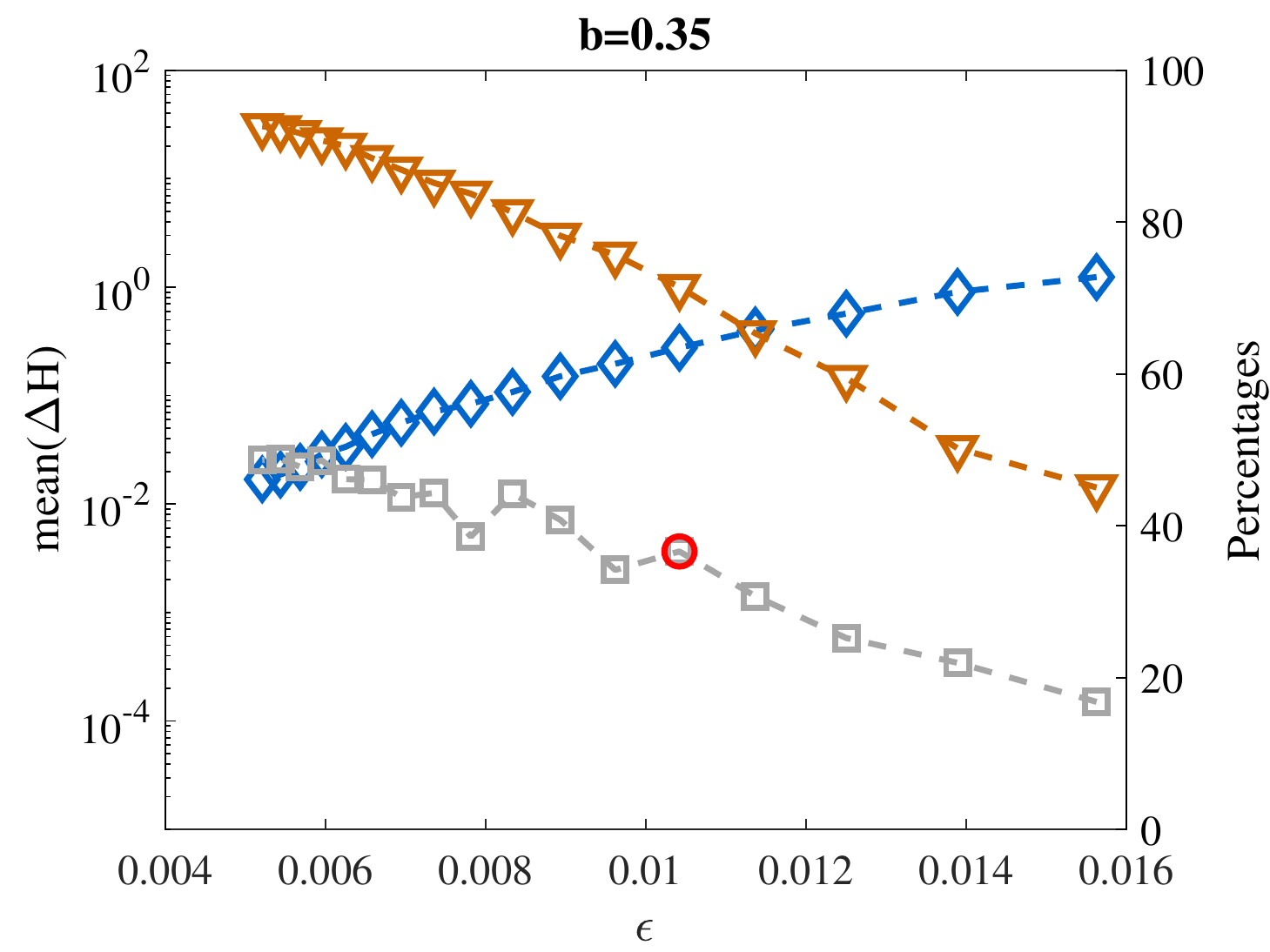} 
\\
\includegraphics[width=0.48\hsize]{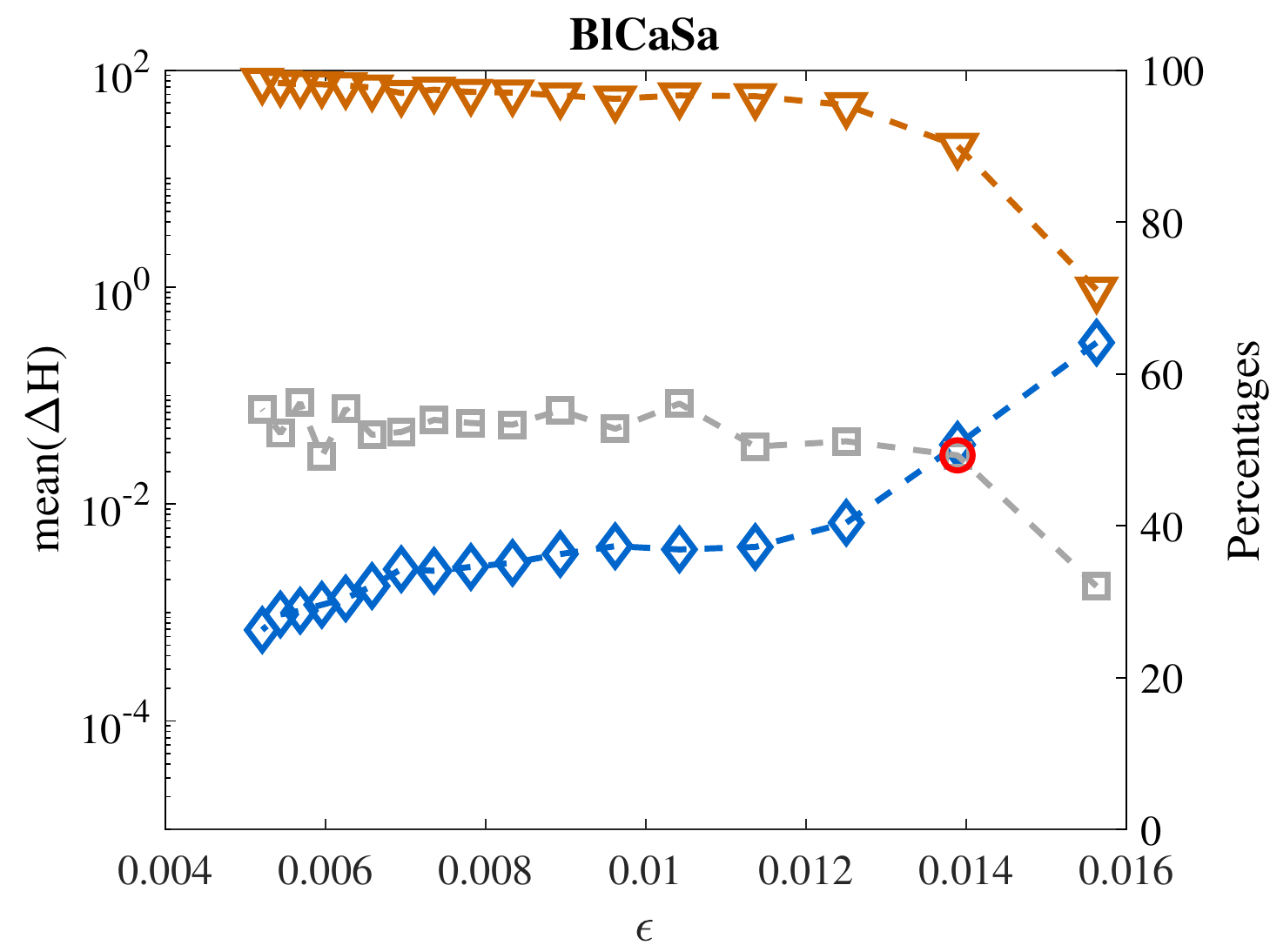}
\quad
\includegraphics[width=0.48\hsize]{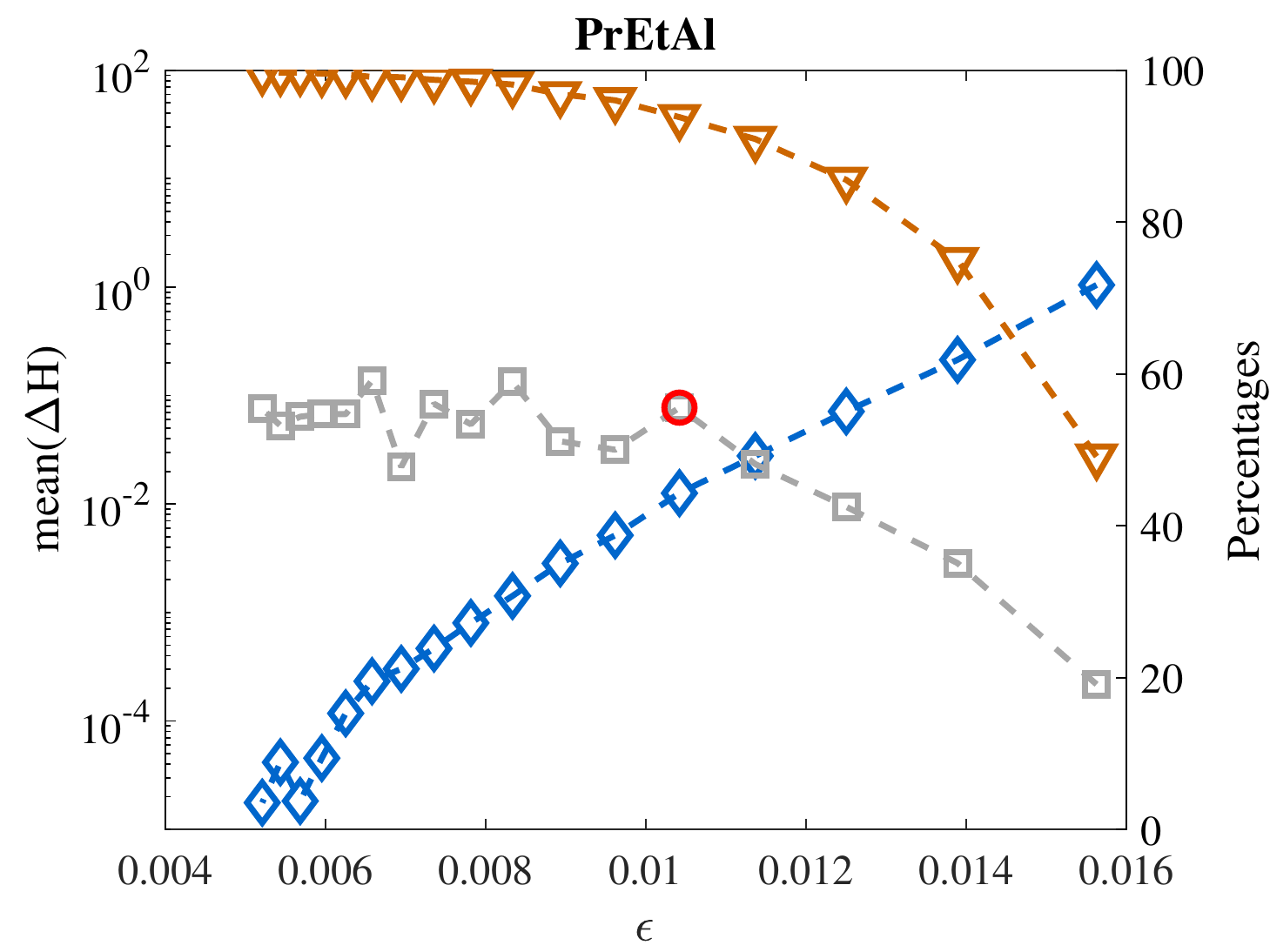}
\\
\includegraphics[width=0.48\hsize]{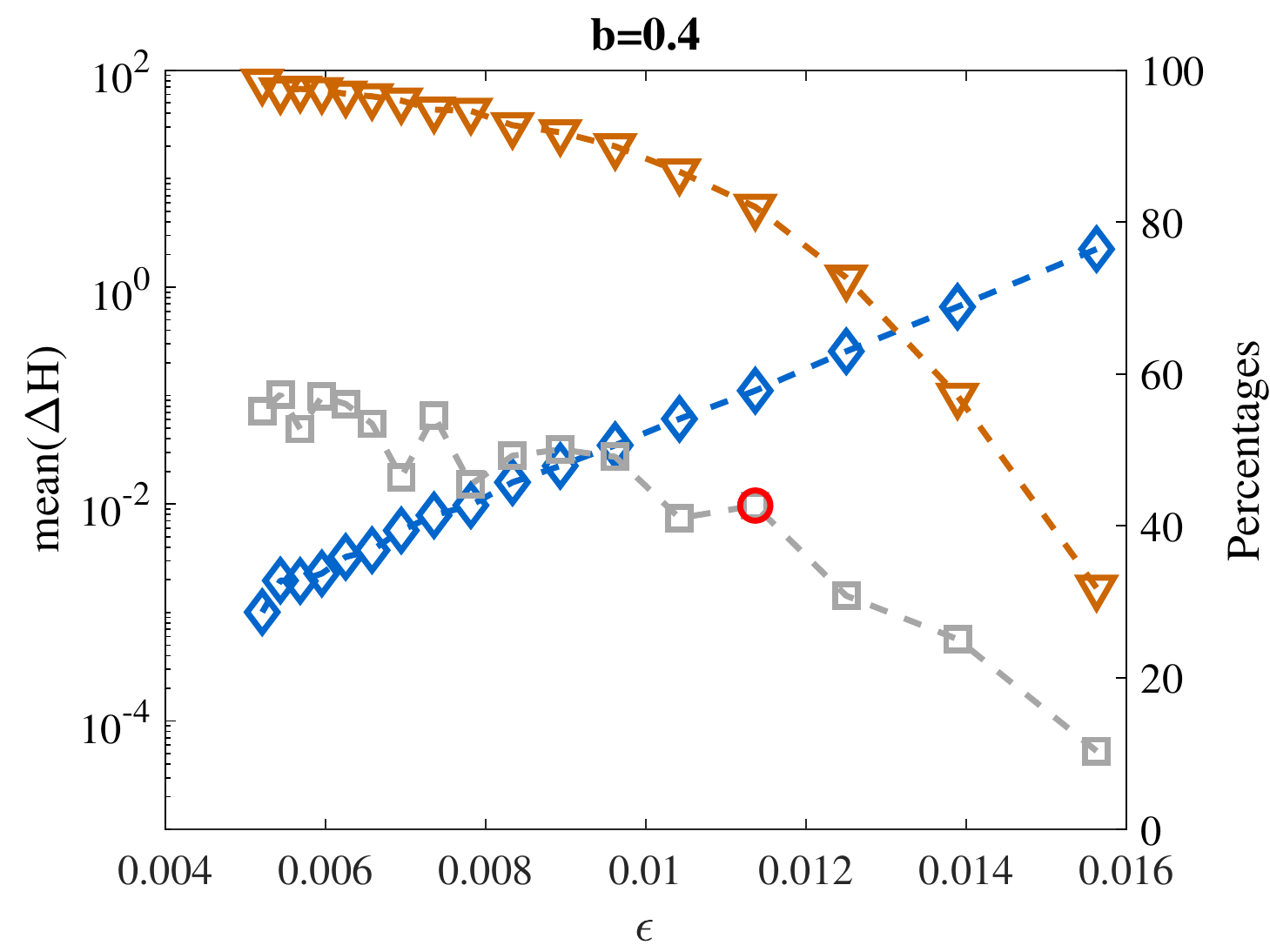}
\quad
\includegraphics[width=0.48\hsize]{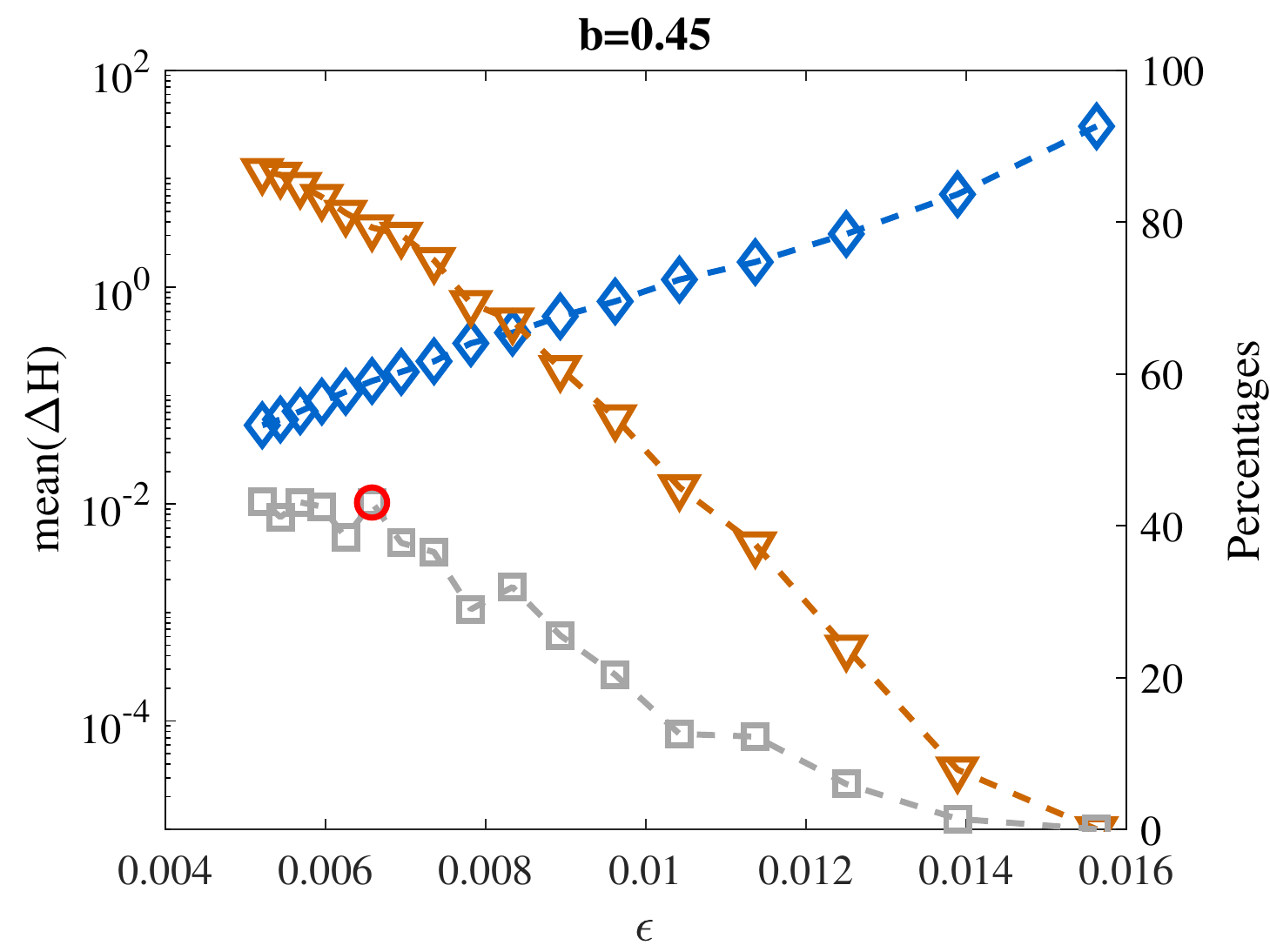}
\caption{Gaussian model, $d=256$. ESS percentage for \(\theta_1\) (squares), acceptance percentage (triangles) and the mean of $\Delta H $ (diamonds) as functions of the step-length $\epsilon$, for different values of the integrator parameter $b$, when $\tau_{end}=5$. The red circle highlights for each method the most efficient run as measured
 by the highest ESS per time-step.}
\label{Gaussian256}
\end{center}
\end{figure}

\begin{figure}
\begin{center}
\includegraphics[width=0.48\hsize]{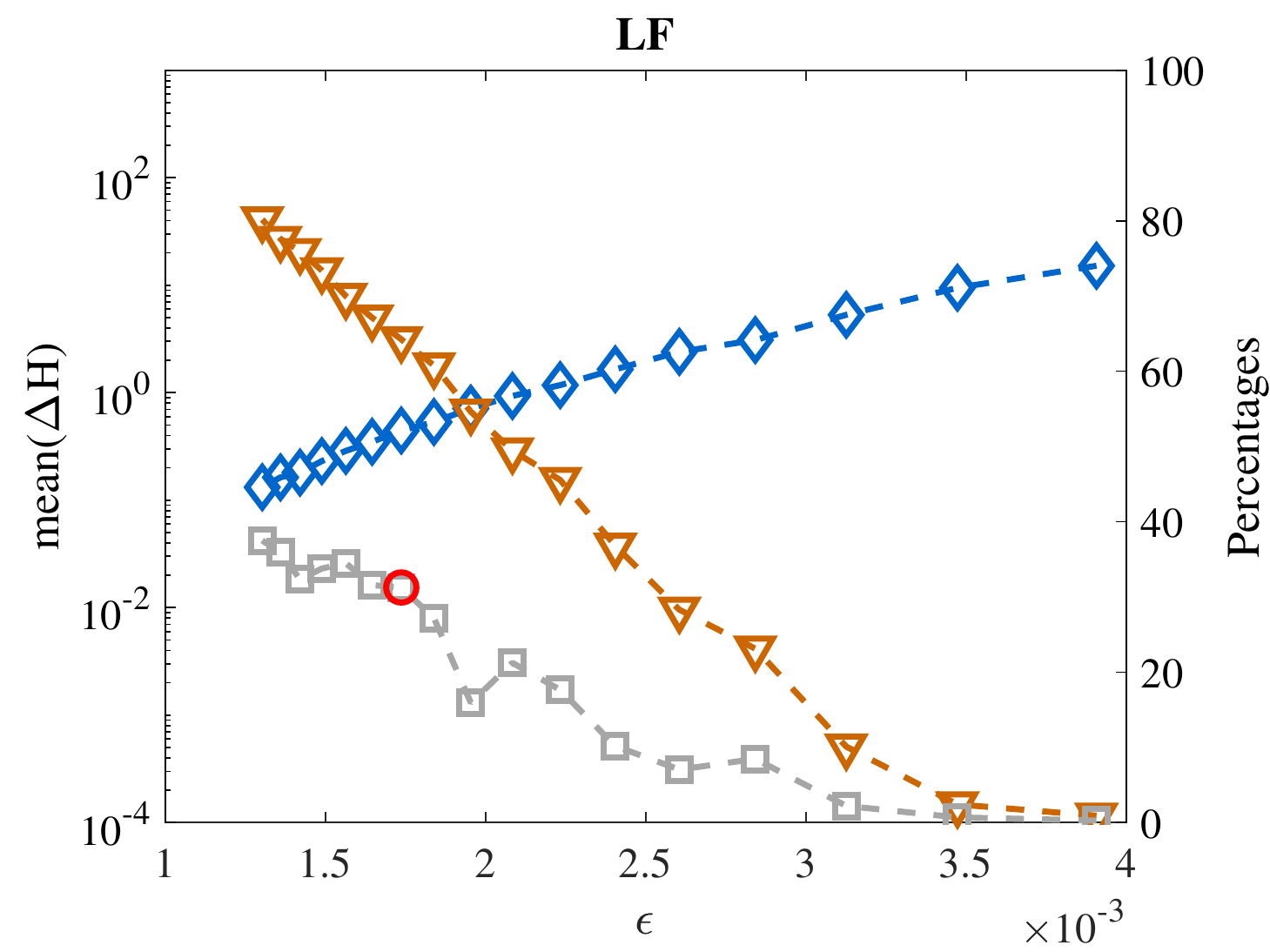}
\quad
\includegraphics[width=0.48\hsize]{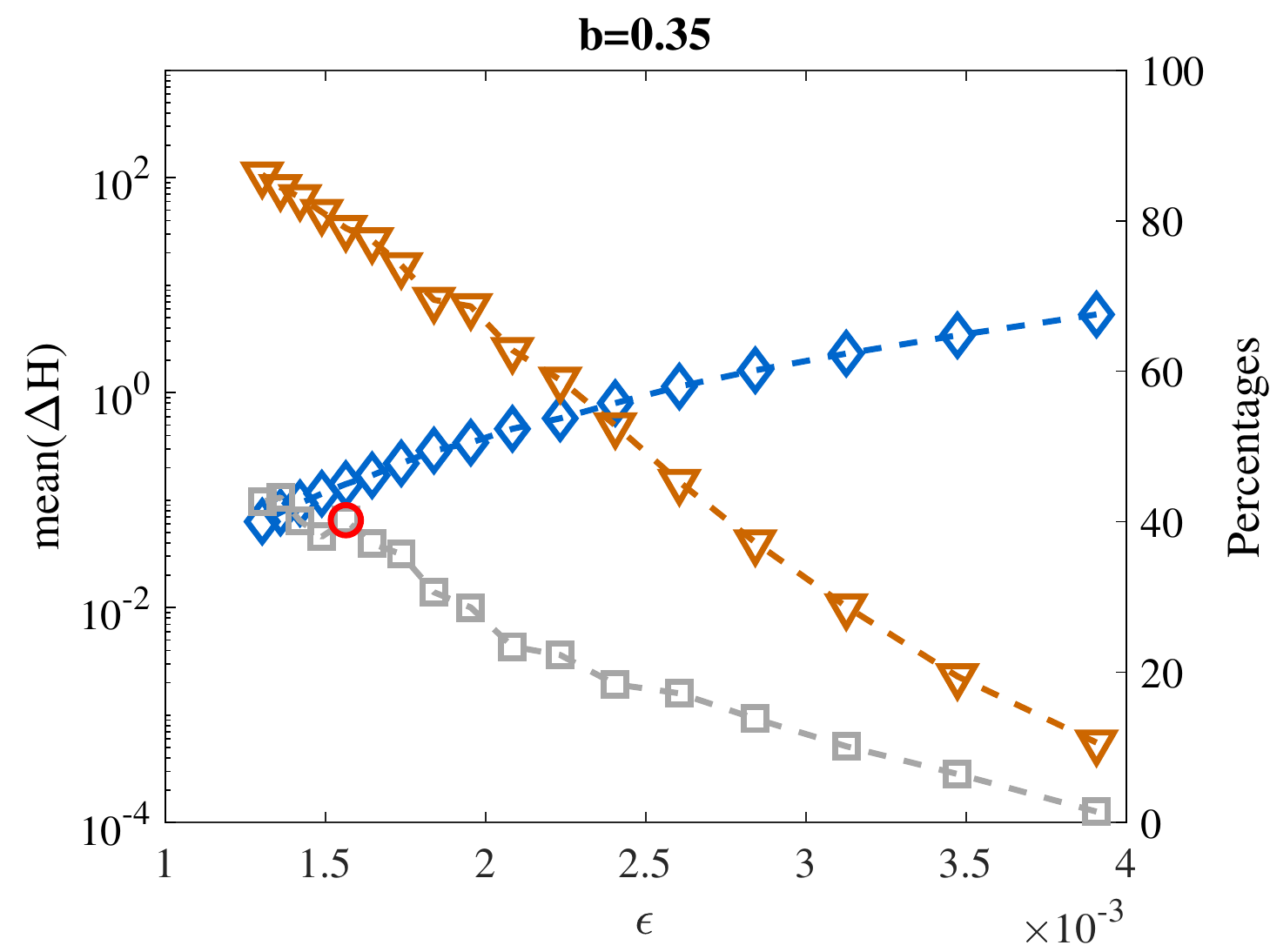} 
\\
\includegraphics[width=0.48\hsize]{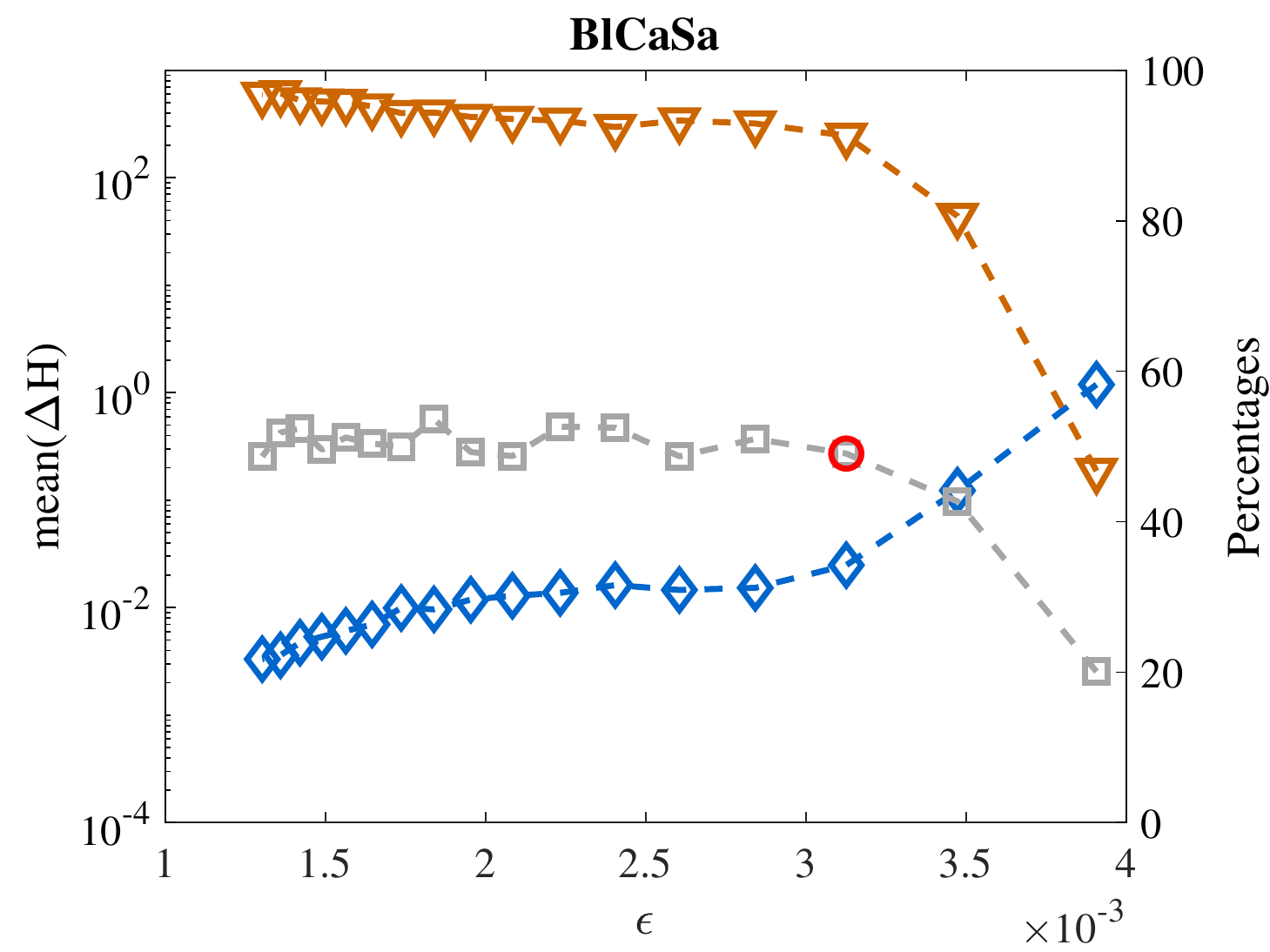}
\quad
\includegraphics[width=0.48\hsize]{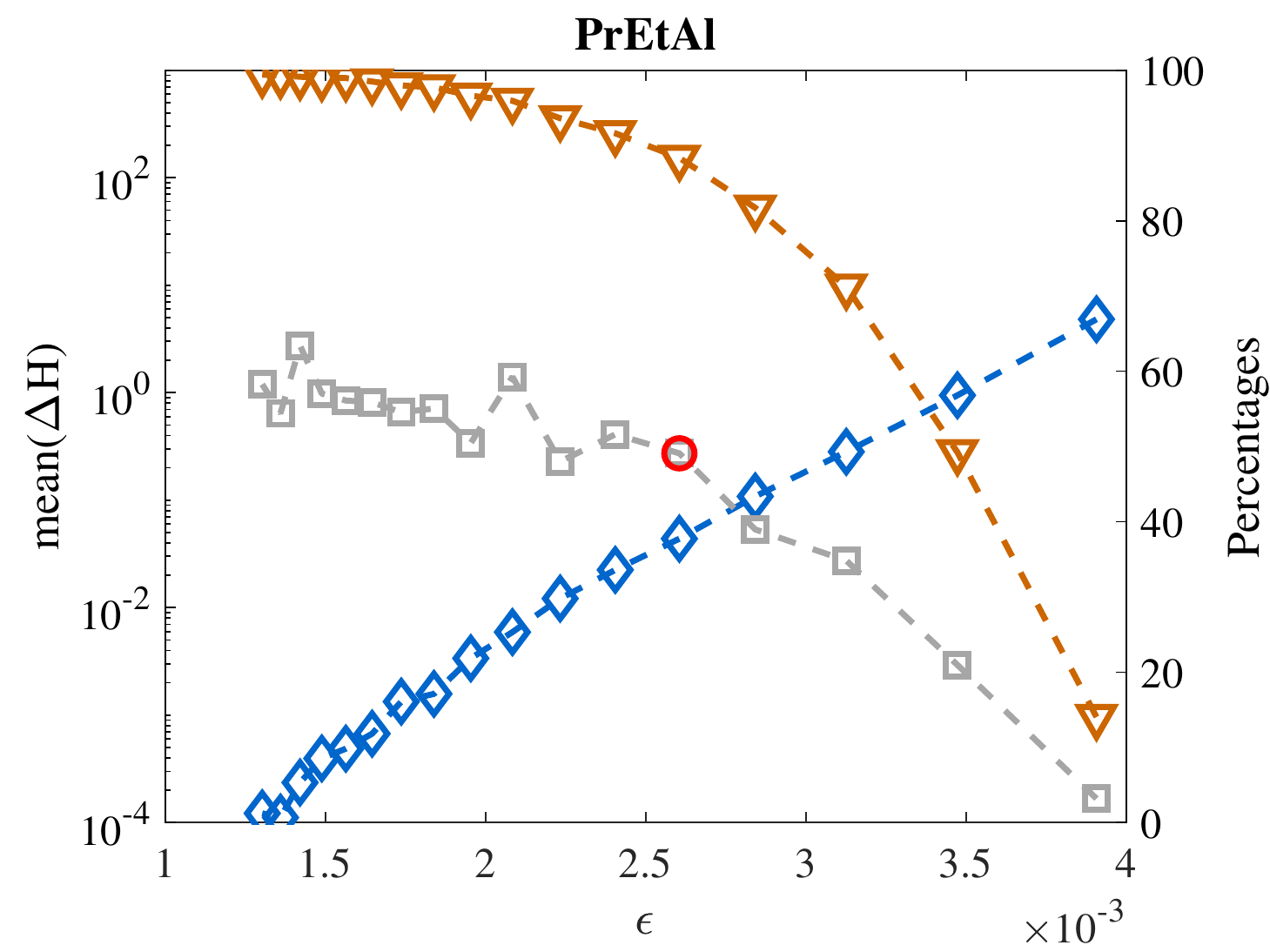}
\\
\includegraphics[width=0.48\hsize]{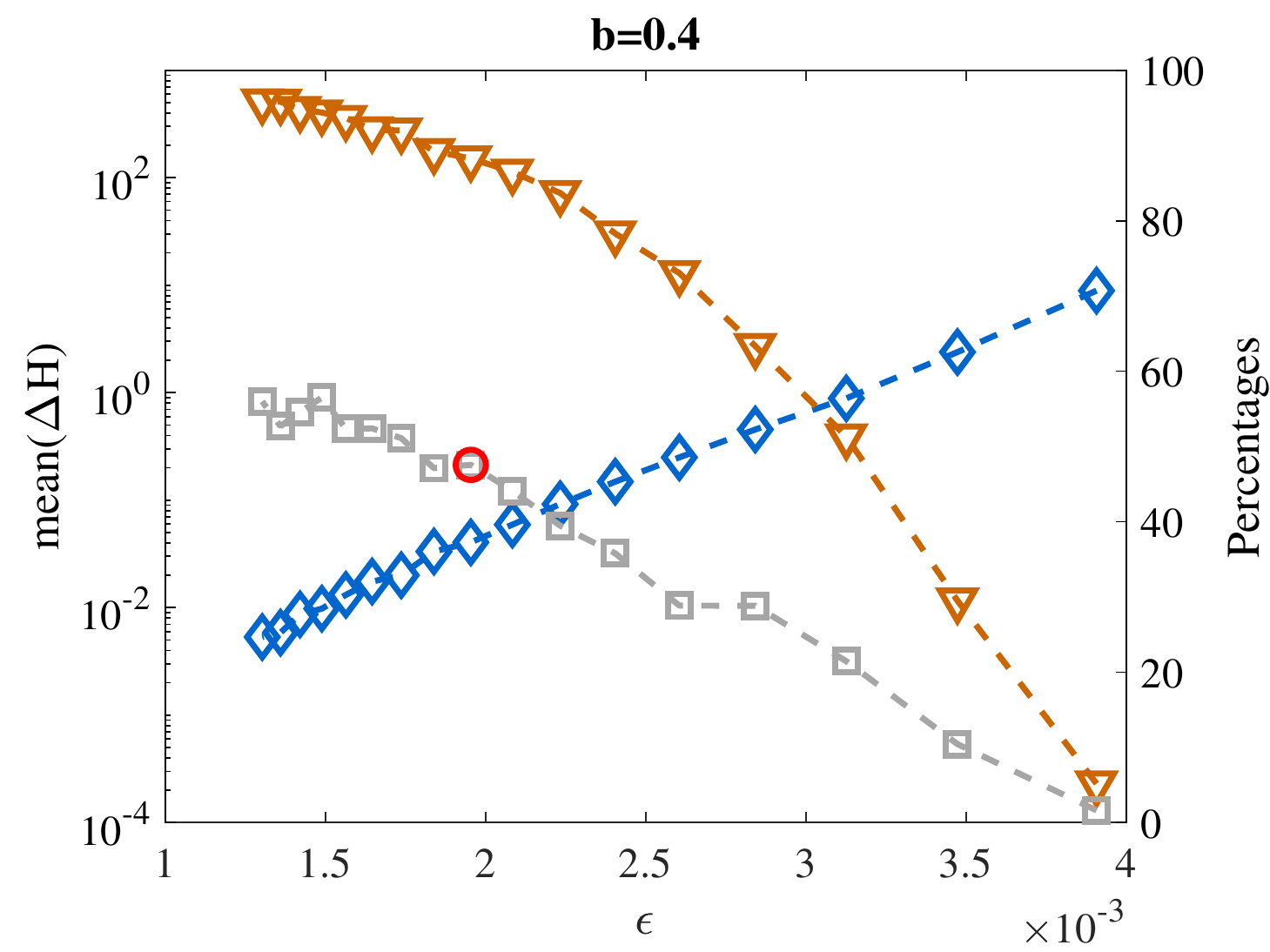}
\quad
\includegraphics[width=0.48\hsize]{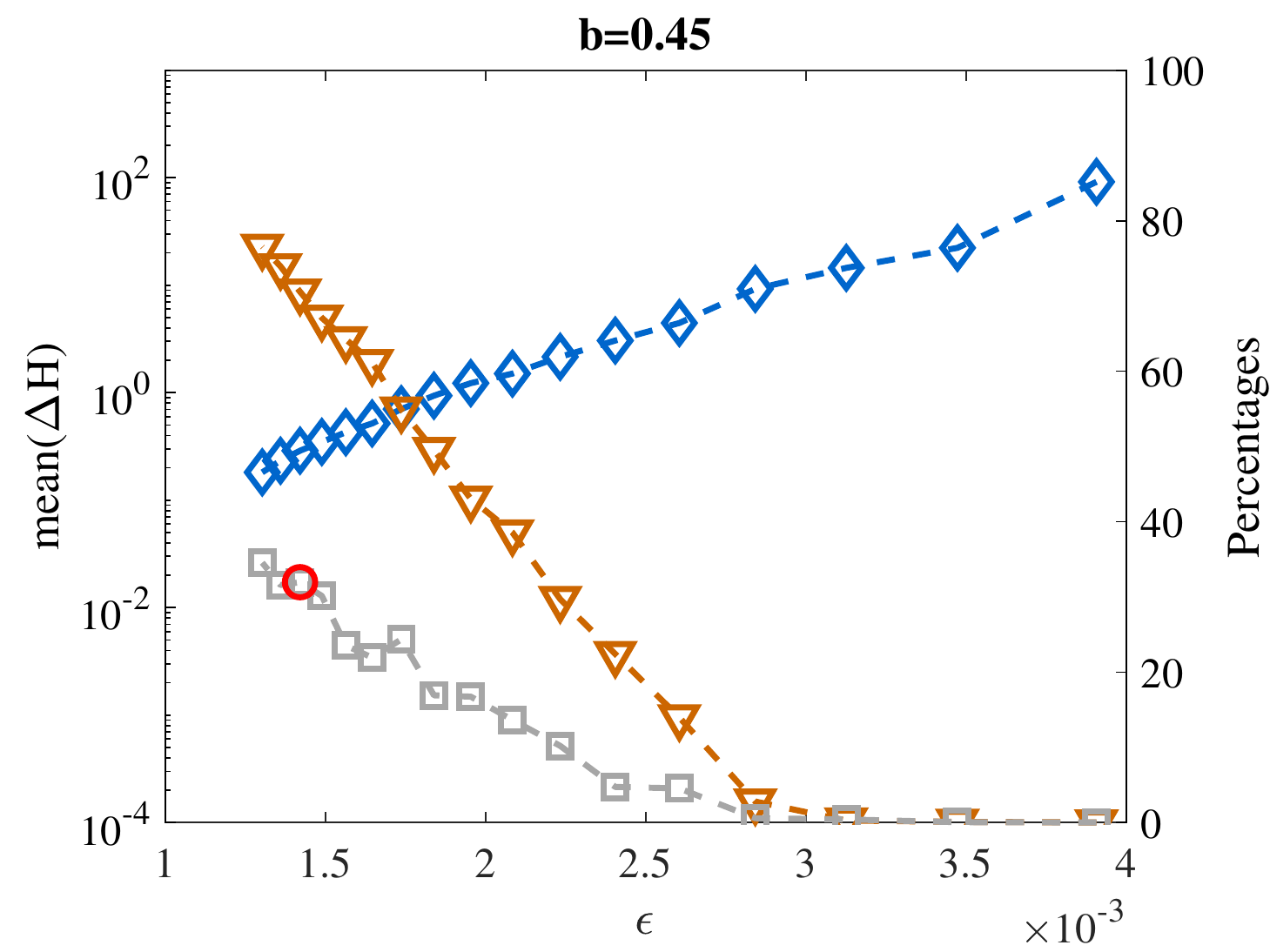}
\caption{Model Gaussian, $d=1024$. ESS percentage for \(\theta_1\) (squares), acceptance percentage (triangles) and the mean of $\Delta H$ (diamonds) as functions of the step-length $\epsilon$, for different values of the integrator parameter $b$, when $\tau_{end}=5$. The red circle highlights for each method the most efficient run as measured by the
 highest ESS per time-step.}
\label{Gaussian1024}
\end{center}
\end{figure}

For the different methods
 we have plotted in Figures \ref{Gaussian256} (\(d=256\)) and \ref{Gaussian1024} (\(d=1024\)), as functions of the step-size \(\epsilon\), the mean of the changes in \(H\) (diamonds), the acceptance percentage (triangles),
 and the ESS  of \(\theta_1\) as a percentage of the  number \(N\) of samples in the Markov chain (squares). Note that we use a logarithmic scale for  the energy increments (left vertical axis) and a linear scale for acceptance rate
 and ESS (right vertical axis).


 We look first at the  mean of \(\Delta H\) as a function of \(\epsilon\) for the different integrators.  When comparing Figure~\ref{Gaussian256} with Figure~\ref{Gaussian1024}, note that both the vertical and horizontal scales are different; \(d=1024\) uses smaller values of \(\epsilon\) and yet shows larger values of the mean energy error.

The differences between BlCaSa, PrEtAl and \(b=0.4\) are remarkable in both figures if we take into account
that these three integrators have values of \(b\) very close to each other.\footnote{When running BlCaSa and
PrEtAl it is important not to round the values of \(b\) given here and to compute accurately the value of
\(a\) from \eqref{eq:constraint}.}
 PrEtAl was designed by making \(\Delta H\) small in the limit \(\epsilon\downarrow 0\) for Gaussian targets
  and we see that indeed the graph of \(\Delta H\) as a function of \(\epsilon\) has a larger slope
   for this method than for the other five integrators. Even though, for \(\epsilon\) small,
   PrEtAl  delivers very small energy errors, this does not seem to be particularly advantageous
    in HMC sampling, because those  errors happen where the acceptance rate is close to \(100\%\),
    i.e.\ not in the range of values of \(\epsilon\) that make sense on efficiency grounds. On the other hand,  BlCaSa was derived so as to get, for Gaussian targets, small energy errors over a range of values of \(\epsilon\); this is borne out in both figures where for this method the variation of \(\Delta H\) with \(\epsilon\) shows a kind of plateau, not present in any of the other five schemes.

In both figures, for each \(\epsilon\), the values of the mean of \(\Delta H\) for each of the integrators
\(b=0.35\), BlCaSa, PrEtAl and \(b=0.4\) are all below the corresponding value for \(b=1/3\). Since, for a
given \(\epsilon\), all integrators have the same computational effort, this implies that  each of those four
methods improves on LF as it gives less error for a given computational cost.

The mean  acceptance as a function of \(\epsilon\) has a behaviour that mirrors that of the mean of \(\Delta
H\): smaller mean energy errors lead to larger mean acceptance rates as expected. For \(\epsilon\) small,
PrEtAl has the highest acceptance rate, but again this is not advantageous for our sampling purposes. In
Figure~\ref{Gaussian256}, BlCaSa shows acceptance rates above \(\approx 70\%\) for the coarsest
\(\epsilon\approx 1.6\times 10^{-2}\); to get that level of acceptance with LF one needs to reduce
\(\epsilon\) to \(\approx 1\times 10^{-2}\). The advantage of BlCaSa over LF in that respect is even more
marked in Figure~\ref{Gaussian1024}.

\begin{figure}[p]
\includegraphics[width=0.48\hsize]{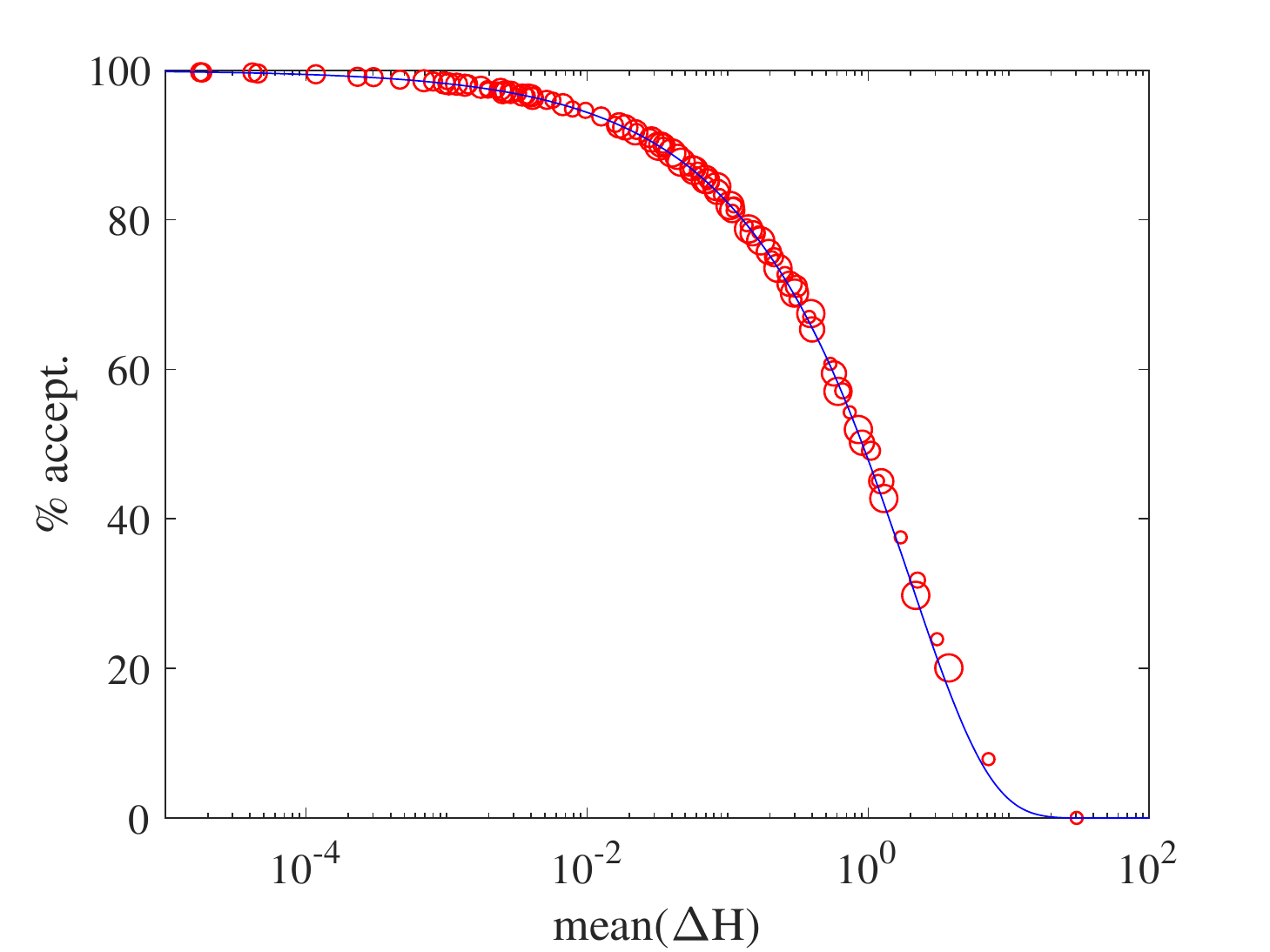} \quad
\includegraphics[width=0.48\hsize]{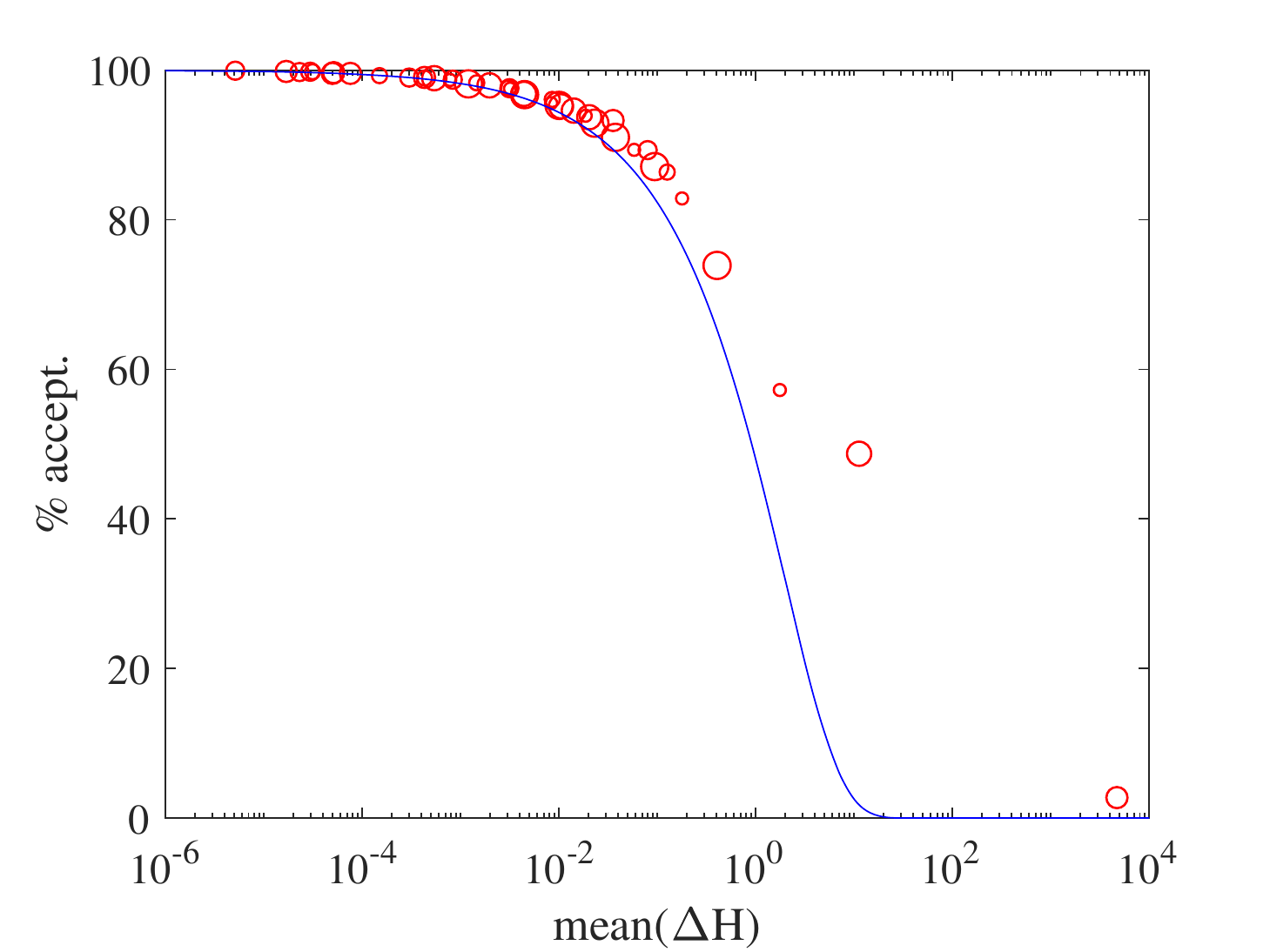}
\caption{Gaussian model, $d=256$ (left), $d=4$ (right). Acceptance percentage vs.\  mean energy error. Points come from all six integrators and all step-lengths. The diameter of the circles indicates the integrator used. Note that the horizontal scales in both plots are different. The solid curve
corresponds to the central limit theorem in Section 5. For \(d=1024\) (not depicted in the figure) the circles fall  exactly (within plotting accuracy) on the solid line.}
\label{Gaussian_fig4_256_1024}
\end{figure}

To visualize  the relation between the mean \(\Delta H\) and the mean acceptance rate, we have plotted in the left panel of Figure~\ref{Gaussian_fig4_256_1024}  the values corresponding to all our runs for \(d=256\)
(i.e.\ to all six integrators and all values of \(\epsilon\)). All points are almost exactly on a curve, so
that to a given value of the mean energy error corresponds a well-defined mean acceptance rate, independently
of the value of \(b\) and \(\epsilon\).  This behaviour is related to a central limit theorem and will be discussed in Section~\ref{sec:mu}.
The right panel displays results corresponding to the six integrators, \(\tau_{end} = 5\), and
different step-lengths for the target \eqref{eq:gaussianexample} for the small value \(d=4\) where the
central limit theorem cannot be invoked; as distinct from the left panel,  it does not
appear to exist a single smooth curve that contains the results of the different integrators.

Returning to Figure~\ref{Gaussian256} and Figure~\ref{Gaussian1024}, we see that in both of them and for all
six methods, the relative ESS is close to \(50\%\) when \(\epsilon\) is very small. This and the convergence of the integrators show empirically that  if we used exact integration of the Hamiltonian dynamics in this simulation, we would see a relative ESS around
\(50\%\). We therefore should prefer integrators that reach that limiting value of \(\approx 50\%\) with the
least computational work, i.e.\ with the largest \(\epsilon\). In this respect the advantage of BlCaSa over LF
is very noticeable. In each panel of both figures, we have marked with a circle the run with the highest ESS
\emph{per unit computational work}, i.e.\ the run where ESS\(\times\epsilon\) is highest. The location of the
circles reinforces the idea that BlCaSa is the method that may operate with the largest \(\epsilon\), i.e.\
with the least computational effort.

\begin{figure}[p]
\includegraphics[width=0.48\hsize]{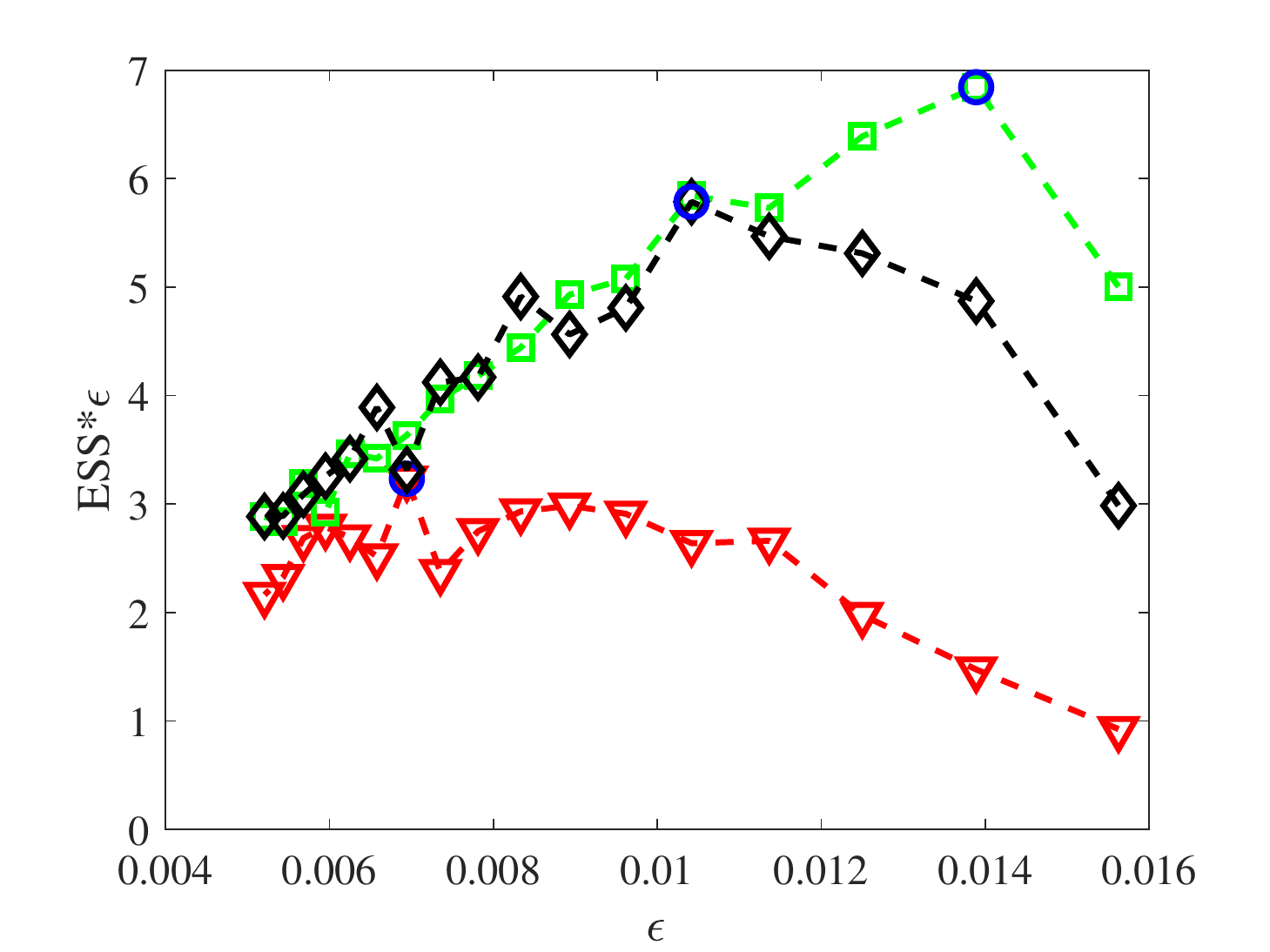} \quad
\includegraphics[width=0.48\hsize]{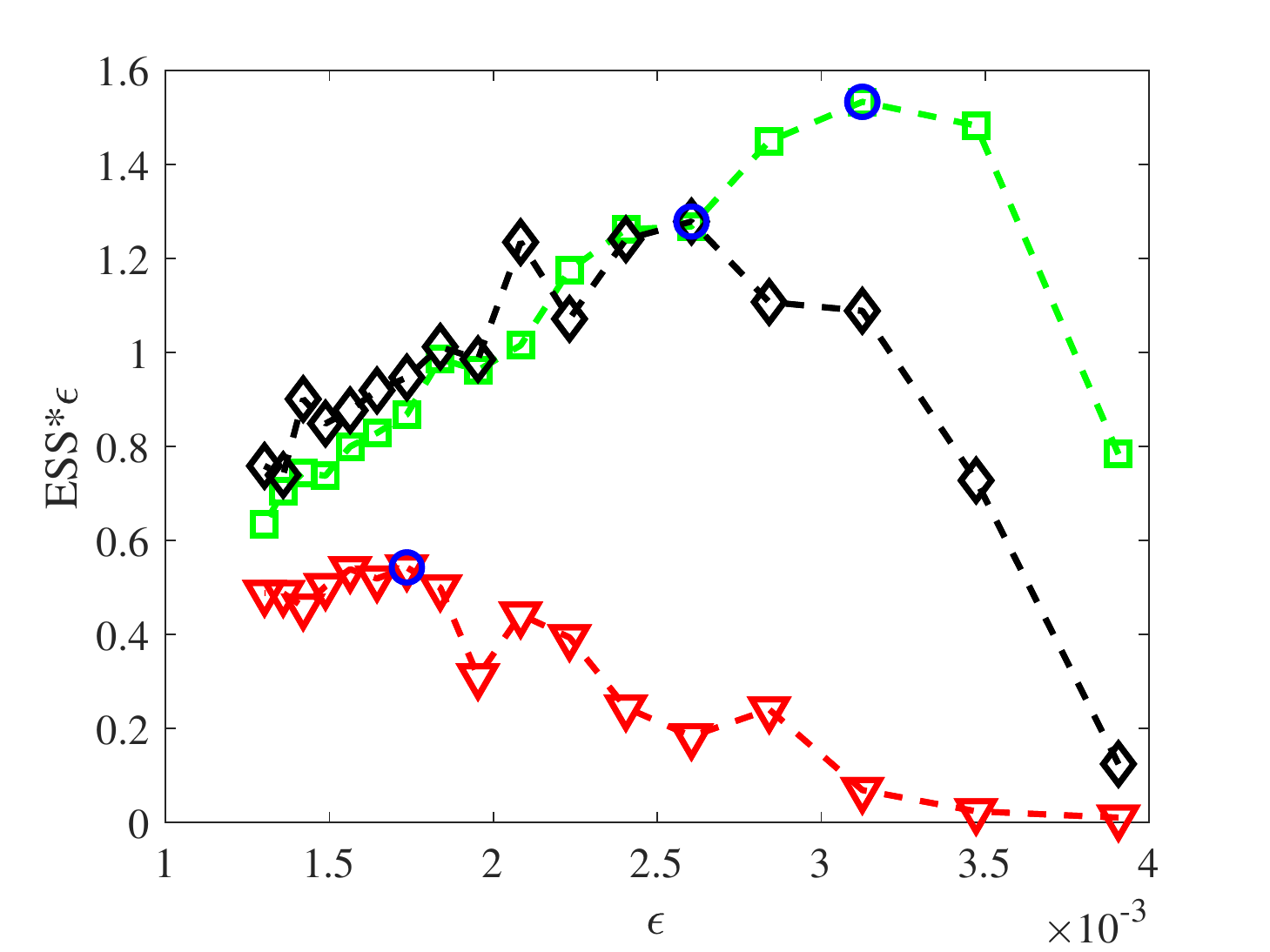}
\caption{Gaussian model, \(d=256\) (left) and \(d=1024\) (right). Comparison of the time integrators BlCaSa (green squares), PrEtAl (black diamonds) and LF (red triangles). ESS per time-step/per unit computational effort  against the step-length $\epsilon$.}
\label{Gaussian_fig3_256}
\end{figure}

In order to better compare the efficiency of the different integration algorithms, we provide  an additional figure,
Figure \ref{Gaussian_fig3_256}. For clarity, only the methods LF, BlCaSa and PrEtAl are considered. In both  panels, the horizontal axis  represents values of \(\epsilon\) and the vertical axis the product  of ESS and \(\epsilon\),  an efficiency metric.
For all step-lengths, the product of ESS and \(\epsilon\), i.e.\ the ESS per unit computational work,
 is clearly smaller
for LF than for the other two methods. Comparing PrEtAl and BlCaSa, we see that for the
largest step-lengths (larger than \(8.0\times 10^{-3}\) when $d=256$ and
larger than \(2.5\times 10^{-3}\) when $d=1024$), BlCaSa shows better values than PrEtAl, but when the step-length decreases the ESS per unit computational work using PrEtAl are slightly higher.  BlCaSa is the most efficient of the three methods, as can be seen by comparing the maxima of the three lines in each panel of both figures.
In both panels a blue circle highlights the most efficient run for each method, which when $d=256$
corresponds to 360 time-steps per leg for BlCaSa (ESS \(=2463\), acceptance rate (AR) \(=90.04\%\)), 480
time-steps per leg for method PrEtAl (ESS \(=2777\), AR \(=93.82\%\)), and 720 time-steps per leg for LF (ESS
\(=2328\), AR \(=81.92\%\)).  Note that the optimal acceptance rate is typically higher than the \(65\%\)
resulting from the analysis of \cite{optimal}. There is no contradiction: while the analysis applies when
\(d\uparrow \infty\) and \(\pi(\theta)\) is a product of identically distributed independent distributions,
 our figures
have a fixed value of \(d\) and here \(\pi(\theta)\) is not of that simple form.

We conclude that in this model problem, for a given computational effort, BlCaSa generates roughly three times
as many effective samples  than LF. In addition the advantage of BlCaSa is more marked as the problem becomes
more challenging.

The information obtained above on the performance of the different integrators when using ESS is essentially the same as when using the mean acceptance rate. For brevity, for the two test problems below we concentrate on the mean acceptance rate. Other metrics lead to the same conclusions both as to the performance of each integrator as \(\epsilon\) varies and as to the relative merits of the integrators.

 \subsection{Log-Gaussian Cox model}
The second test problem \citep{christensen,girolami} is a well-known  high dimensional example of inference in
a log-Gaussian Cox process. A $64 \times 64$ grid is considered in $[0,1] \times [0,1]$ and the random
variables ${\bf X}=X_{i,j}$, $1 \leq i, j \leq 64$, represent the number of points in cell $(i,j)$. These
random variables are conditionally independent and Poisson distributed with means $m \Lambda_{i,j}$, where
$m=1/4096$ is the area of each cell and $\Lambda_{i,j}$, $1 \leq i, j \leq 64$, are unobserved intensities,
which are assumed to be given by $\Lambda_{i,j} = \exp{(Y_{i,j})}$, with ${\bf Y}=Y_{i,j}$, $1 \leq i,j \leq
64$, multivariate Gaussian with mean $\mu$ and covariance matrix $\Sigma$ given by
$$\Sigma_{(i,j),(i',j')} = \sigma^2 \exp{\displaystyle{\left (-\frac{\sqrt{(i-i')^2+(j-j')^2}}{64 \beta}\right )}}.$$
Given the data ${\bf X} = {\bf x}$, the target density  of interest is
$$
\pi({\bf y})\propto
\prod_{i,j=1}^{64} \exp{\left ( x_{i,j} y_{i,j} - m\exp({y_{i,j})}\right )} \exp{\left ( -({\bf y}-\mu {\bf 1})^T
\Sigma^{-1}({\bf y}-\mu {\bf 1})/2\right )}.$$

\begin{figure}
\begin{center}
\includegraphics[width=0.48\hsize]{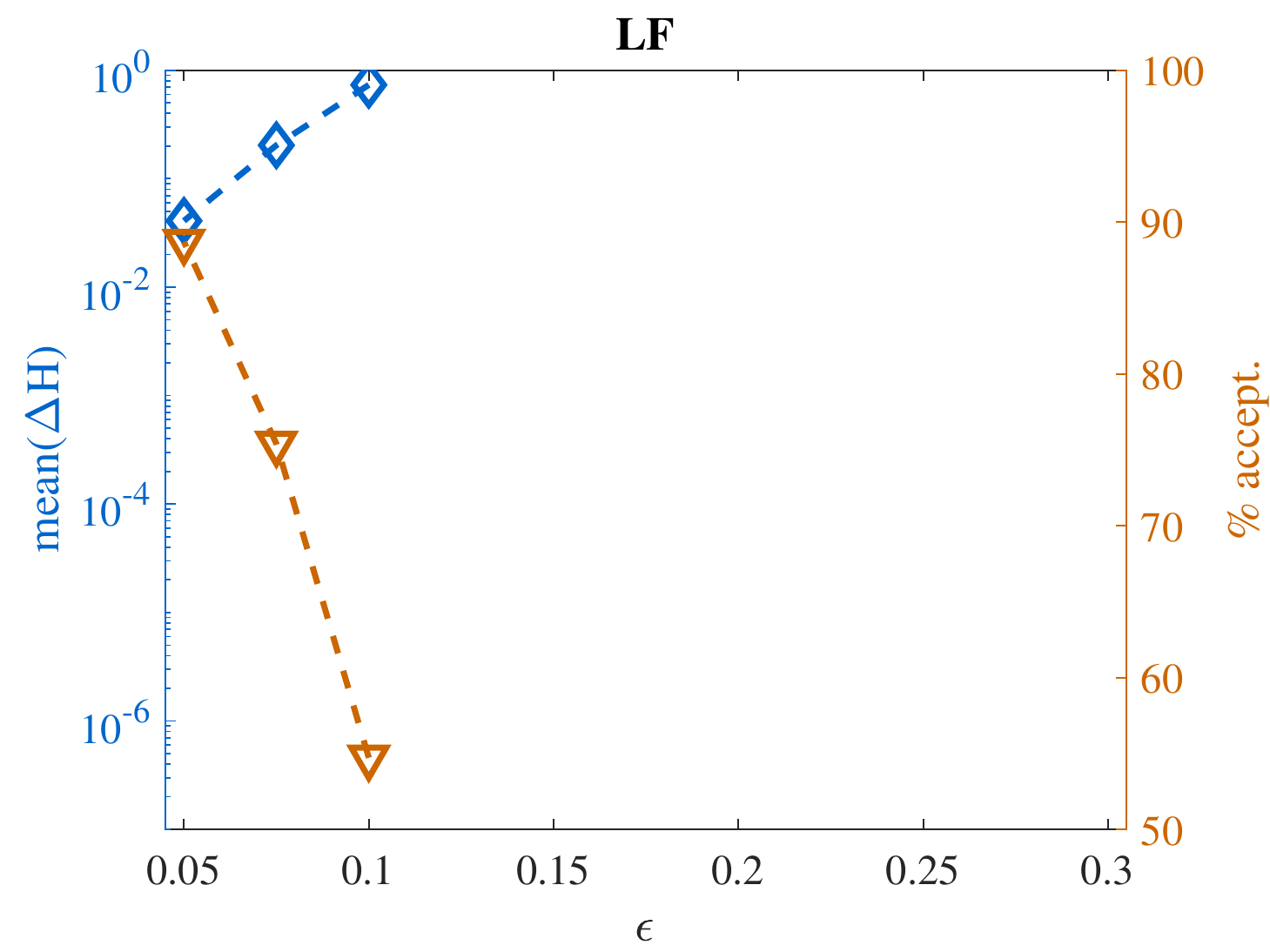} \quad
\includegraphics[width=0.48\hsize]{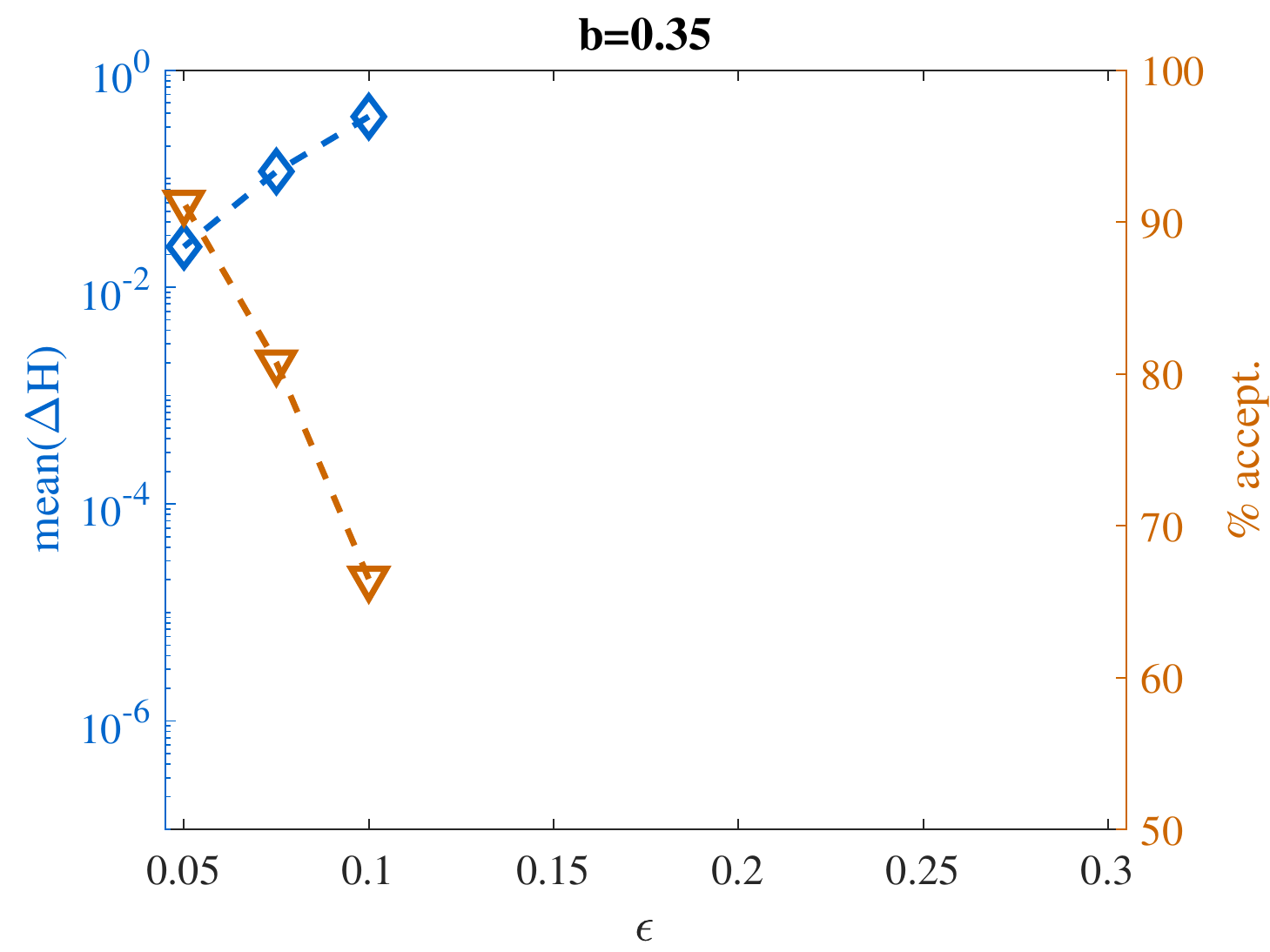}
\\
\includegraphics[width=0.48\hsize]{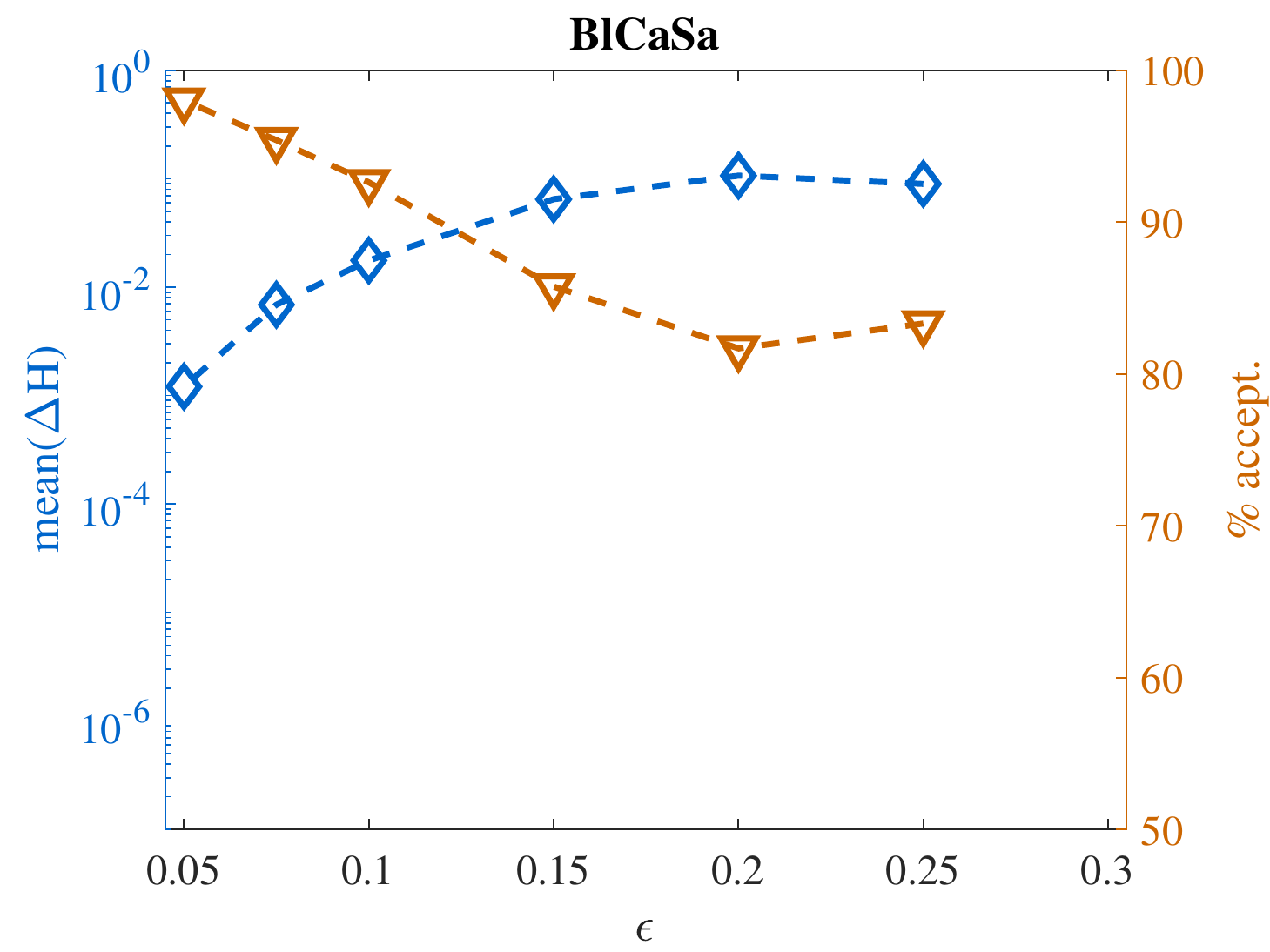} \quad
\includegraphics[width=0.48\hsize]{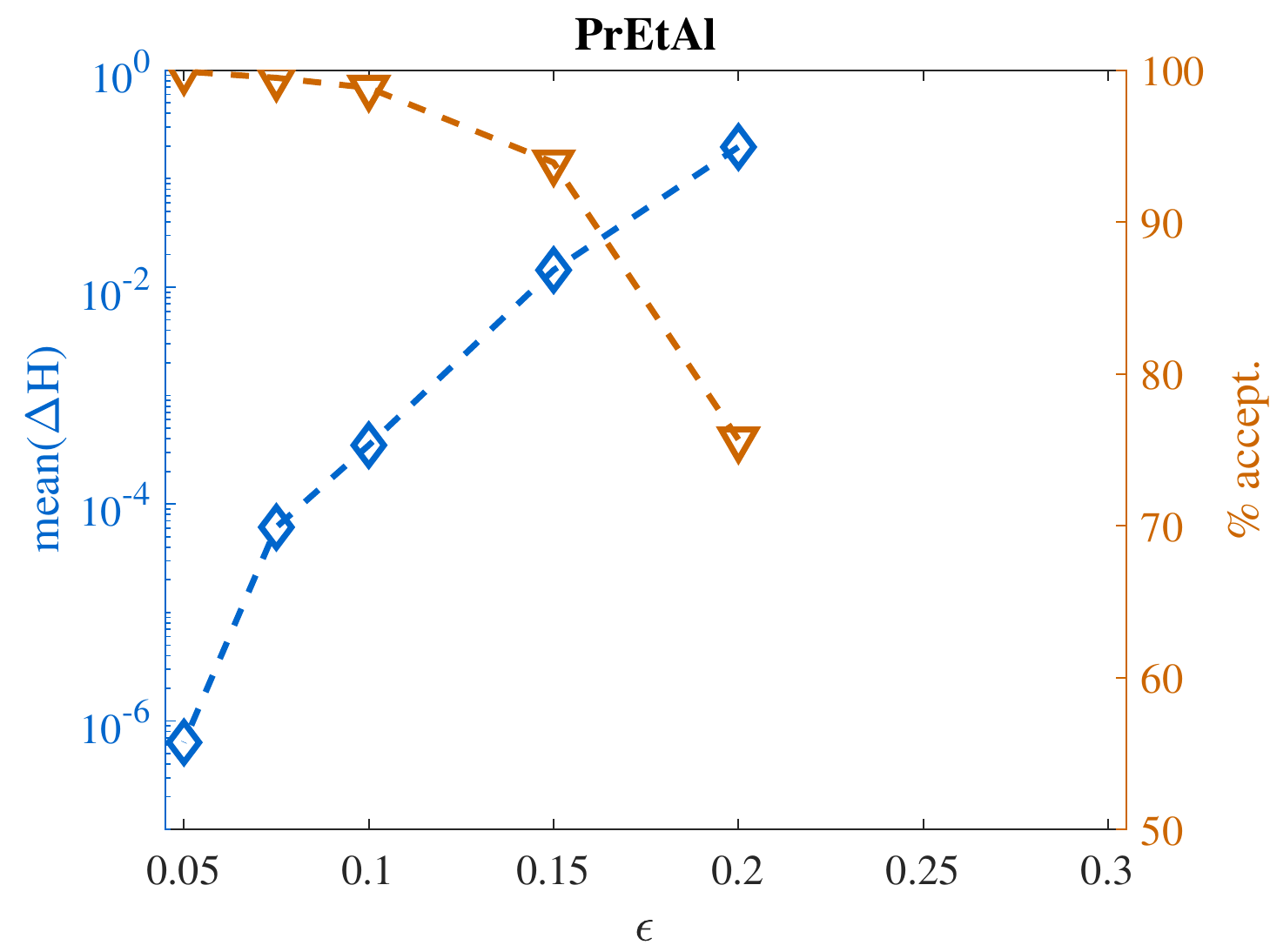}
\\
\includegraphics[width=0.48\hsize]{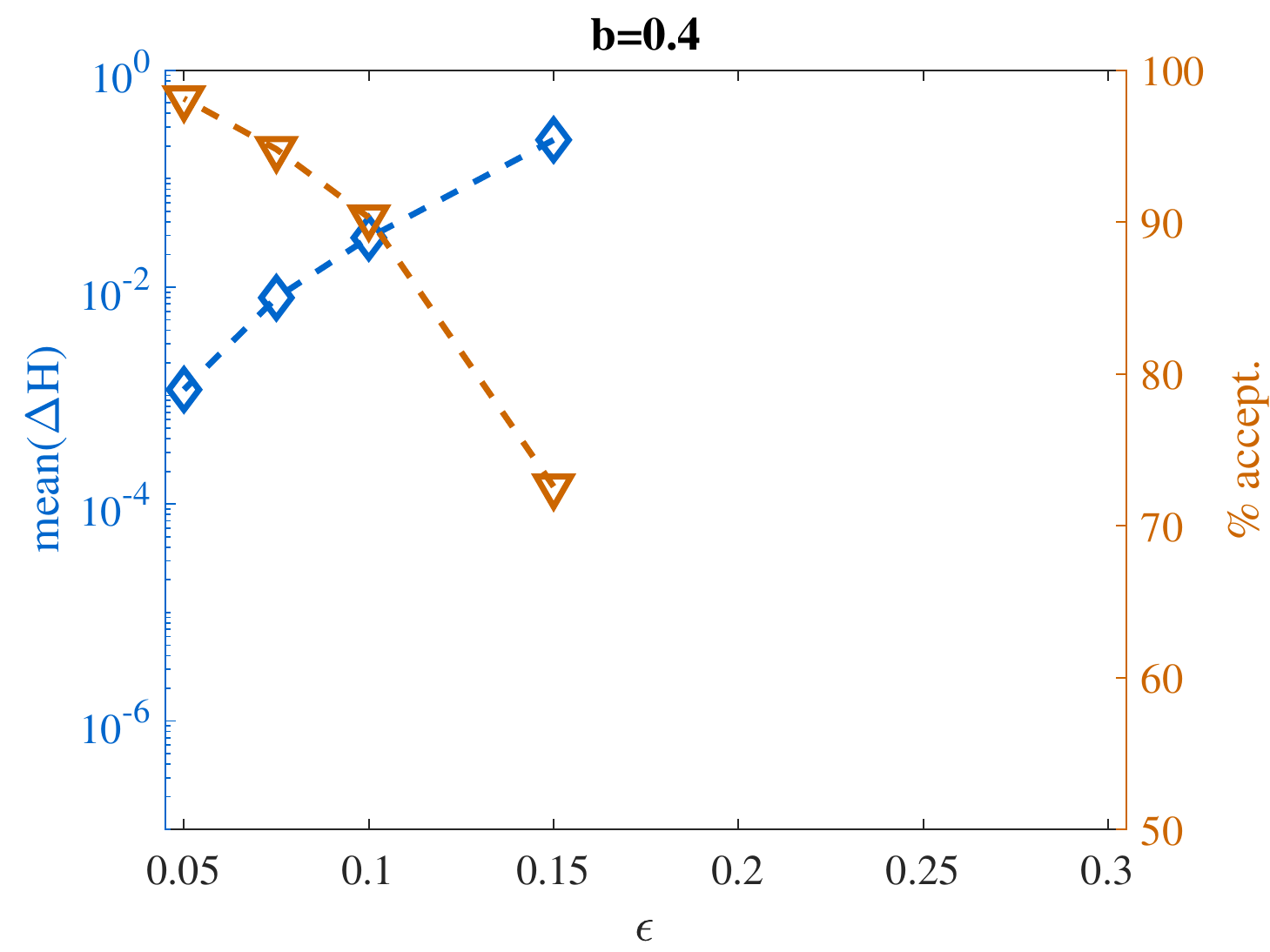} \quad
\includegraphics[width=0.48\hsize]{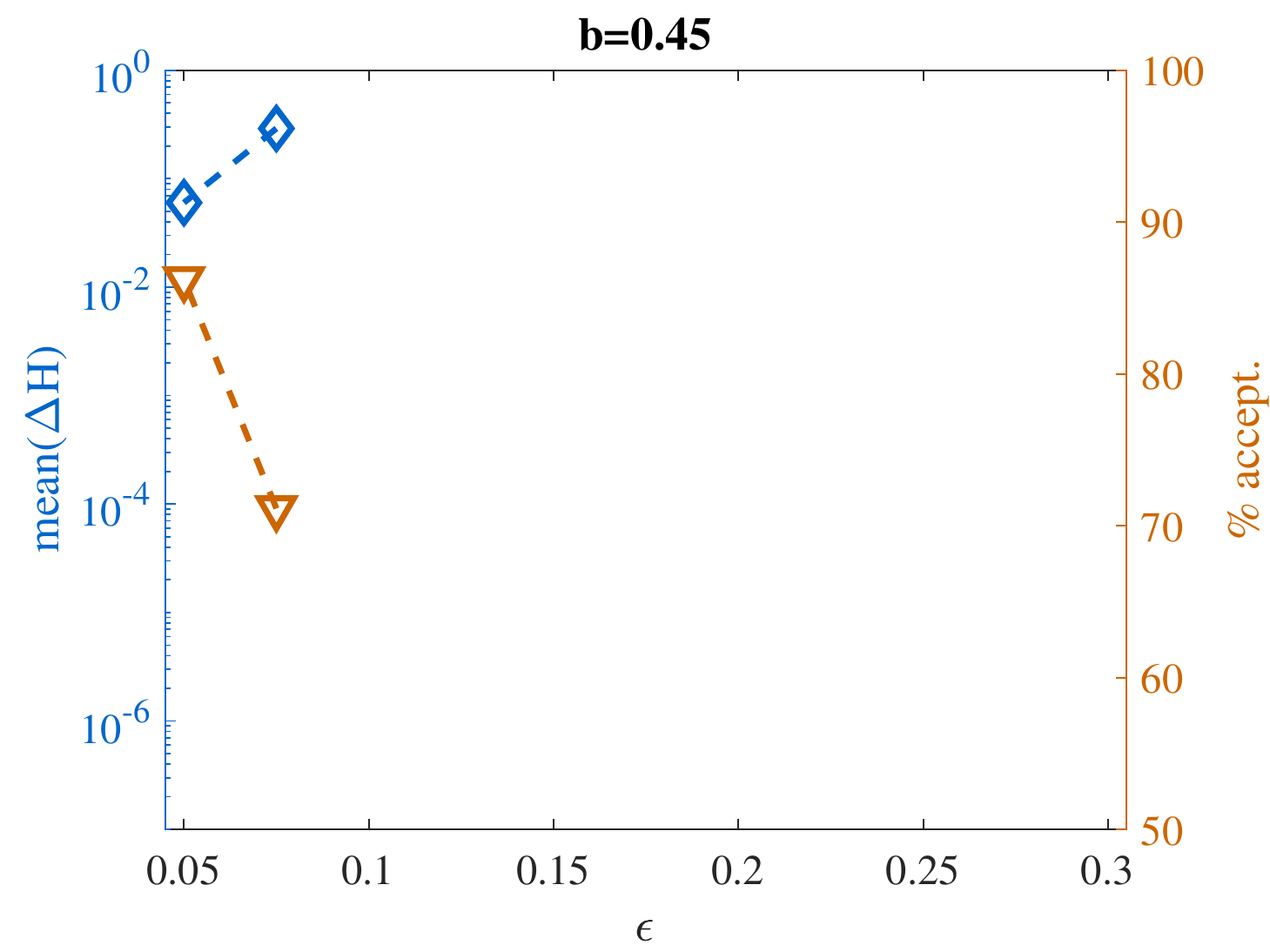}
\caption{Log-Gaussian Cox problem. Acceptance percentage (triangles) and the mean of $\Delta H$ (diamonds) as functions of the
step-length $\epsilon$, for different values of parameter $b$, when $\tau_{end}=3$.}
\label{ini_cond_b}
\end{center}
\end{figure}

\begin{figure}
\begin{center}
\includegraphics[width=0.48\hsize]{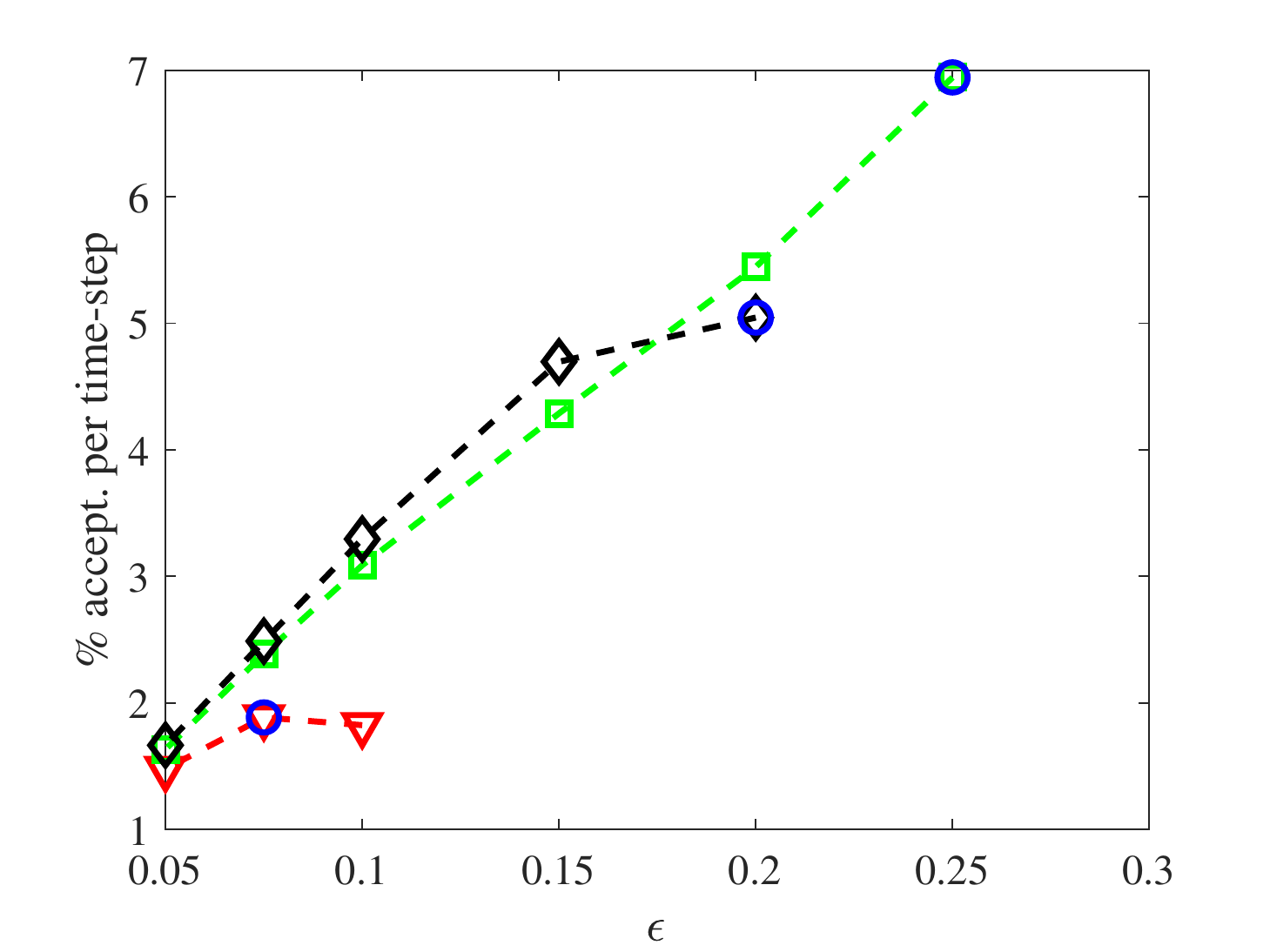}
\caption{Log-Gaussian Cox problem. Acceptance percentage per time-step as a function of $\epsilon$, for PrEtAl (diamonds), BlCaSa (squares) and LF (triangles), when $\tau_{end}=3$.}
\label{Cox_fig5}
\end{center}
\end{figure}

As in \cite{christensen} we have fixed the parameters $\beta=1/33$, $\sigma^2 = 1.91$ and $\mu=\log{126}-\sigma^2/2 \approx 3.881$ and used three alternative starting values. The conclusions obtained as to the merits of the different integrators  are very similar for the three alternative cases and, for this reason, we only report here results corresponding to one of them  (choice (b) in \cite{christensen}). The initial state is randomly constructed for each run in the following way. Given a random vector $\Gamma$ with components \({\cal N}(0,1)\) the starting value ${\bf y}$ is the solution of the nonlinear system ${\bf y} = \mu {\bf 1} + L \Gamma$, where $L$ is the Cholesky factor of the matrix $(\Sigma^{-1} + \mbox{diag}({\bf y}))^{-1}$ which, therefore, depends on ${\bf y}$. This nonlinear system has been solved by fixed-point iteration starting with ${\bf y}^{(0)} = \mu {\bf 1}$, and the iteration has been stopped when $\|{\bf y}^{(n+1)}-{\bf y}^{(n)}\|<10^{-12}$ (which happens after 19 iterations).

The length of each integration leg has been set equal to 3, and the \lq\lq basic\rq\rq\ step-lengths used are
$\epsilon=0.3, 0.25, 0.20, 0.15, 0.1, 0.075, 0.05$. After generating \(1000\)  burn-in samples, we have run
Markov chains with \(5000\) elements, a number that is sufficient for our purposes.

In Figure \ref{ini_cond_b} we have represented the acceptance percentage  and the mean of the increments in energy against the step-size $\epsilon$ for the different values of the parameter $b$. We have omitted data corresponding to either acceptance percentages below $45$ or values of the mean of $\Delta H$ larger than 1.

Comparing the performance of the six methods, we see that for this problem, BlCaSa, in contrast with the other
five integrators,
 operates well for \(\epsilon = 0.25\). PrEtAl gives very high acceptance
 rates for small values of \(\epsilon\), but this is not very relevant for the reason given
 in the preceding test example.
The other four integrators, including LF, require very small step-lengths to generate
 a reasonable number of accepted proposals; their performance is rather poor.

The energy error in PrEtAl decays by more than five orders of magnitude
 when reducing \(\epsilon\) from \(0.2\) to \(0.05\); this is remarkable
 because the additional  order of energy accuracy of this integrator  as \(\epsilon\downarrow 0\),
 as compared with the other five,  only holds analytically for \emph{Gaussian targets} \citep{predescu}.
 Similarly BlCaSa shows here the same kind of plateau in the mean energy error we observed
 in the Gaussian model.

In Figure \ref{Cox_fig5} we have plotted the acceptance percentage per step
as a function of the step-length $\epsilon$ for the two methods with the largest
 acceptance rates, PrEtAl (diamonds) and BlCaSa (squares) in addition to LF (triangles).
 The blue circles in the figure highlight the most efficient run for each method.  When the
 computational cost is taken into account, BlCaSa is the most efficient
  of the methods being compared, by a factor of more than three with respect to LF.

\subsection{Alkane molecule}
\label{alkane}
The third test problem comes from molecular simulations and was chosen to consider multimodal
 targets very far away from Gaussian models. The target is the Boltzmann distribution \(\pi(\theta)\propto \exp(-V(\theta))\), where the vector \(\theta\) collects the cartesian coordinates of the different atoms in the molecule and \(V(\theta)\) is the potential energy and, accordingly,
 in  \eqref{eq:new1}--\eqref{eq:new7}, \(\cal L(\theta)\) has to be replaced by \(-V(\theta)\).
We have considered  the simulation of an alkane molecule described in detail in \cite{cances}, where it is used to compare HMC (with leapfrog) and many alternative samplers; HMC shows a good performance and has no difficulty in identifying the different modes (stable configurations of the molecule).    The expression for the potential energy is  complicated, as it contains contributions related to bond lengths, bond angles, dihedral angles, electrostatic forces and Van der Waals forces. In our numerical experiments molecules with 5, 9 or 15 carbon atoms have been used, leading to Hamiltonian systems with 15, 27 and 45 degrees of freedom (the hydrogen atoms are not simulated).  The temperature parameter in
\cite{cances} is set to 1 here.

 As \cite{campos}, we set the length of each integration leg as $\tau_{end}=0.48$, with
 \lq\lq basic\rq\rq\ step-length \(\epsilon= 0.08, 0.06, 0.048, 0.04,\) \( 0.03\), corresponding
  to $6, 8, 10, 12$ and $16$ steps per integration leg.  In all runs,
   after generating \(200\) samples for burn-in, we consider Markov chains with 1000 samples;
   the results provided are obtained by averaging over 20 instances of the experiments. In all cases,
   the initial condition \(\theta^{(0)}\) for the chain is the equilibrium with minimum potential energy.

\begin{figure}
\begin{center}
\includegraphics[width=0.48\hsize]{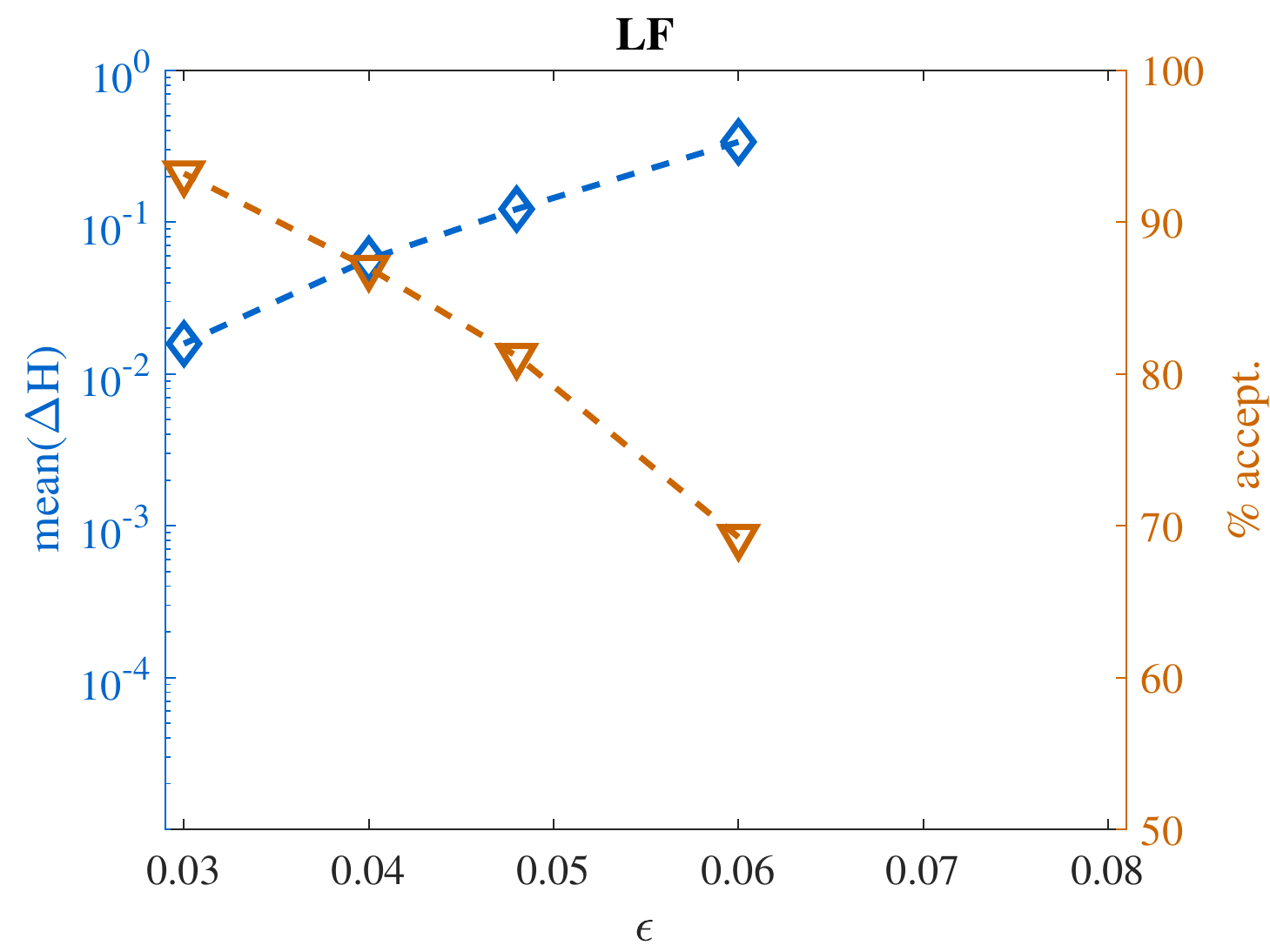} \quad \includegraphics[width=0.48\hsize]{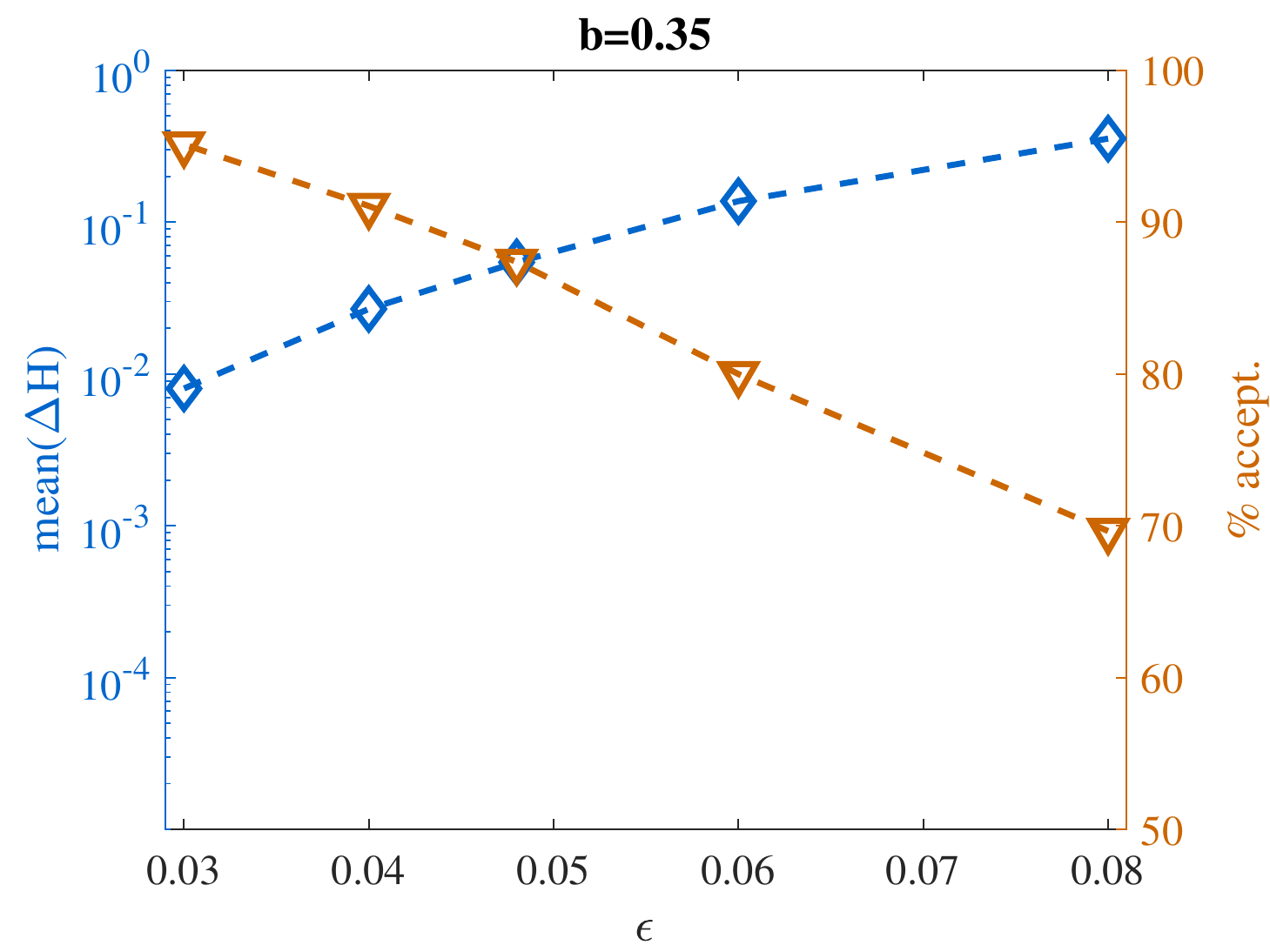}
\\
\includegraphics[width=0.48\hsize]{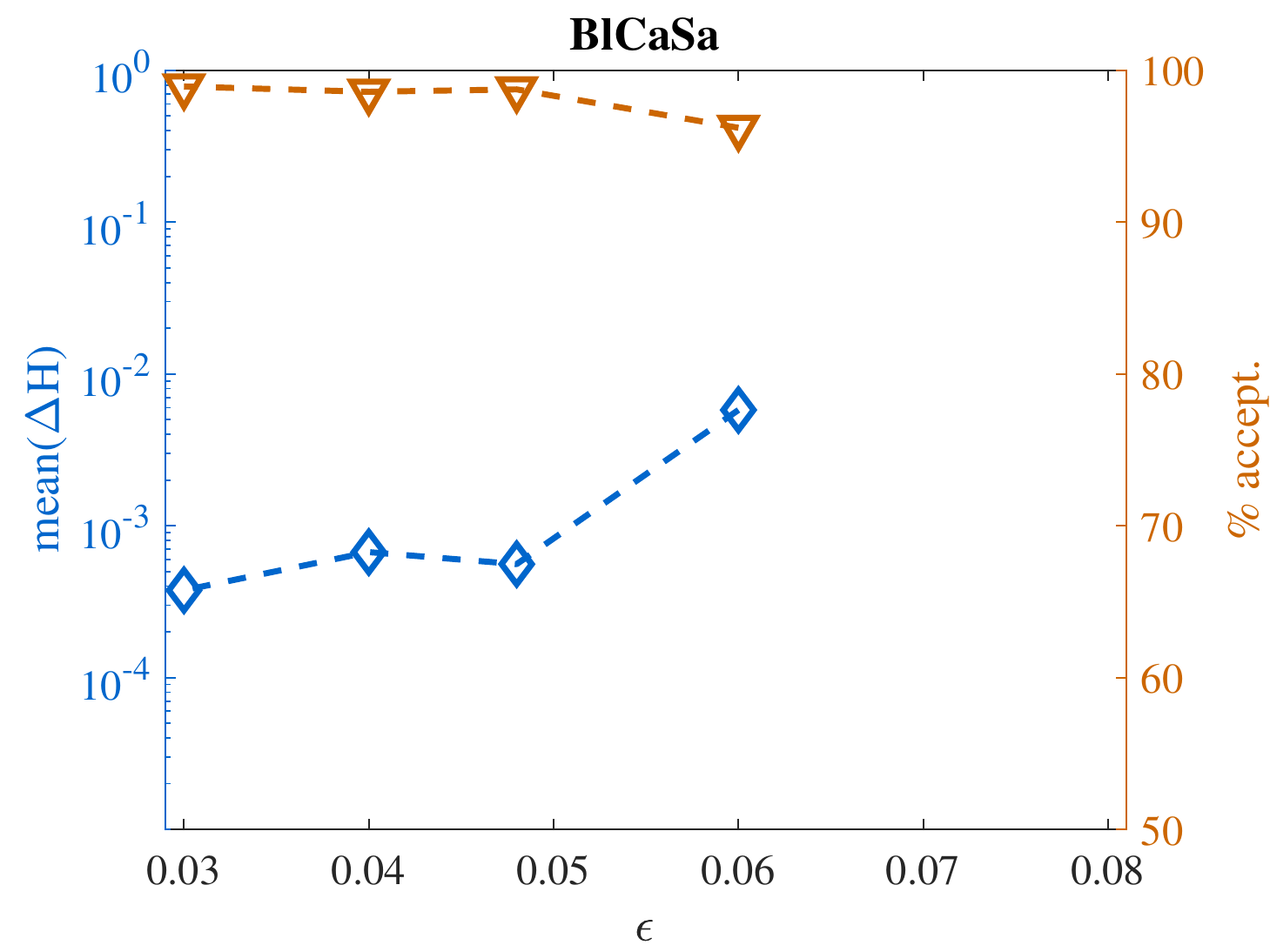}\quad
\includegraphics[width=0.48\hsize]{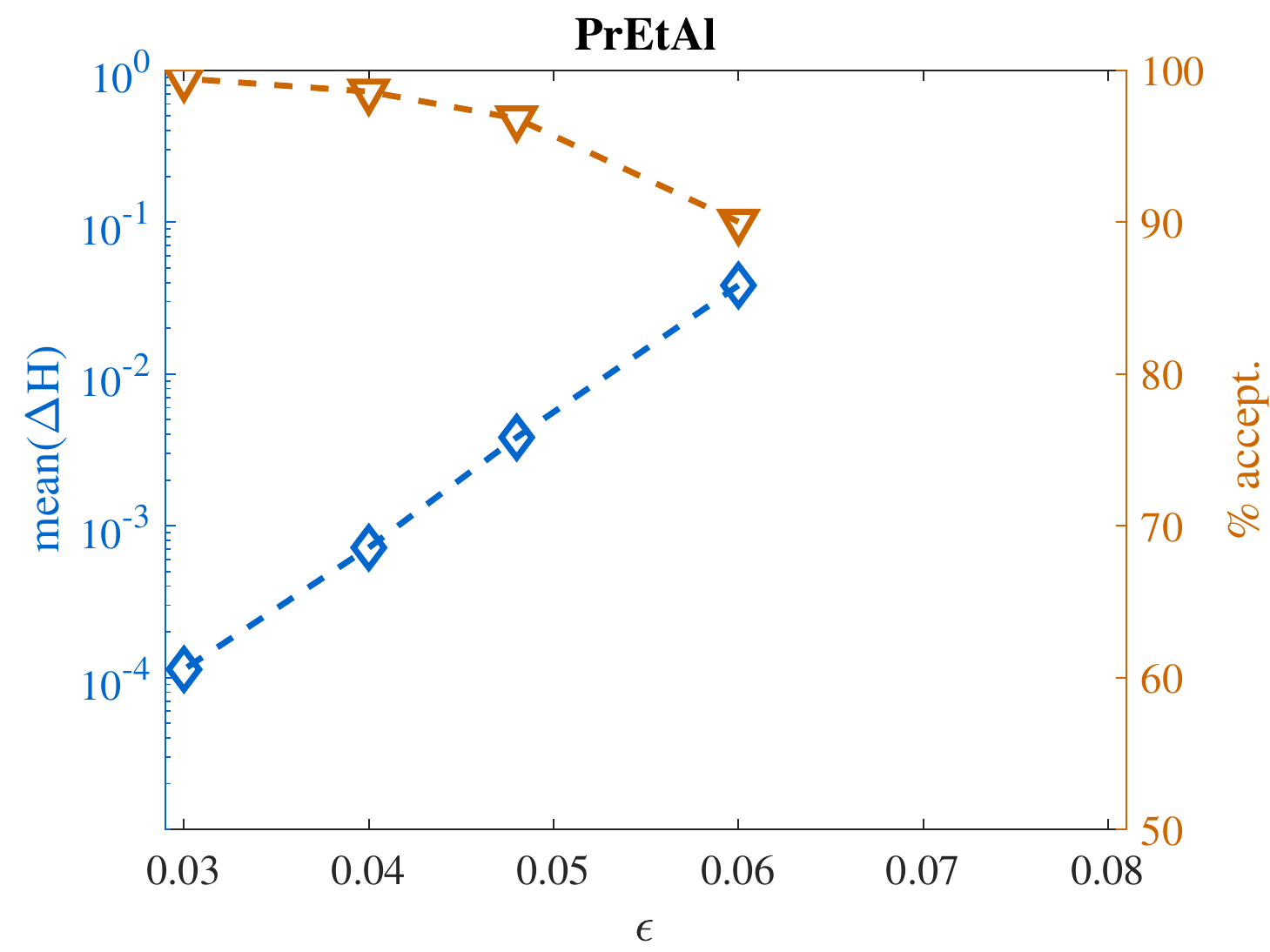}
\\
\includegraphics[width=0.48\hsize]{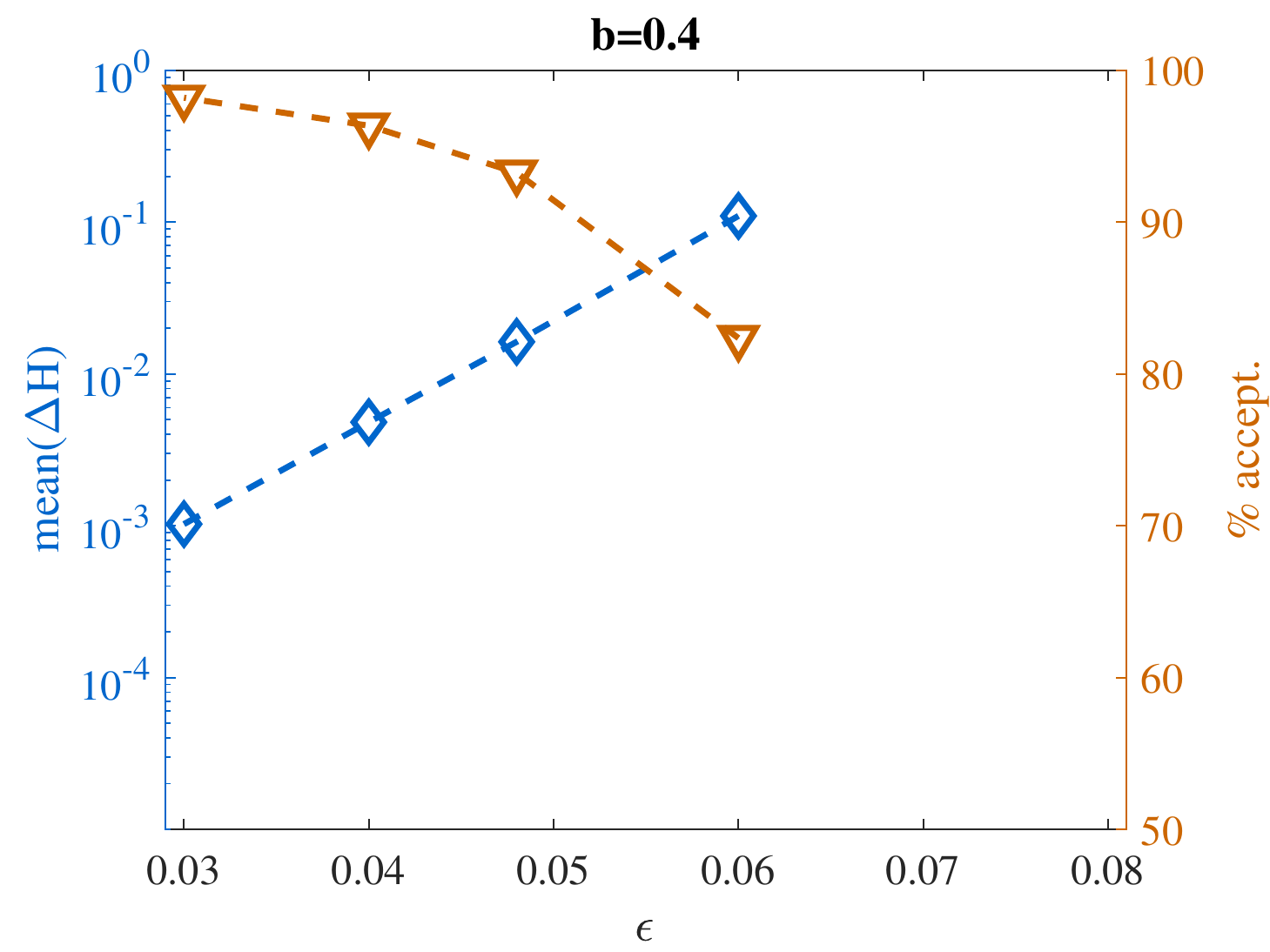}\quad
\includegraphics[width=0.48\hsize]{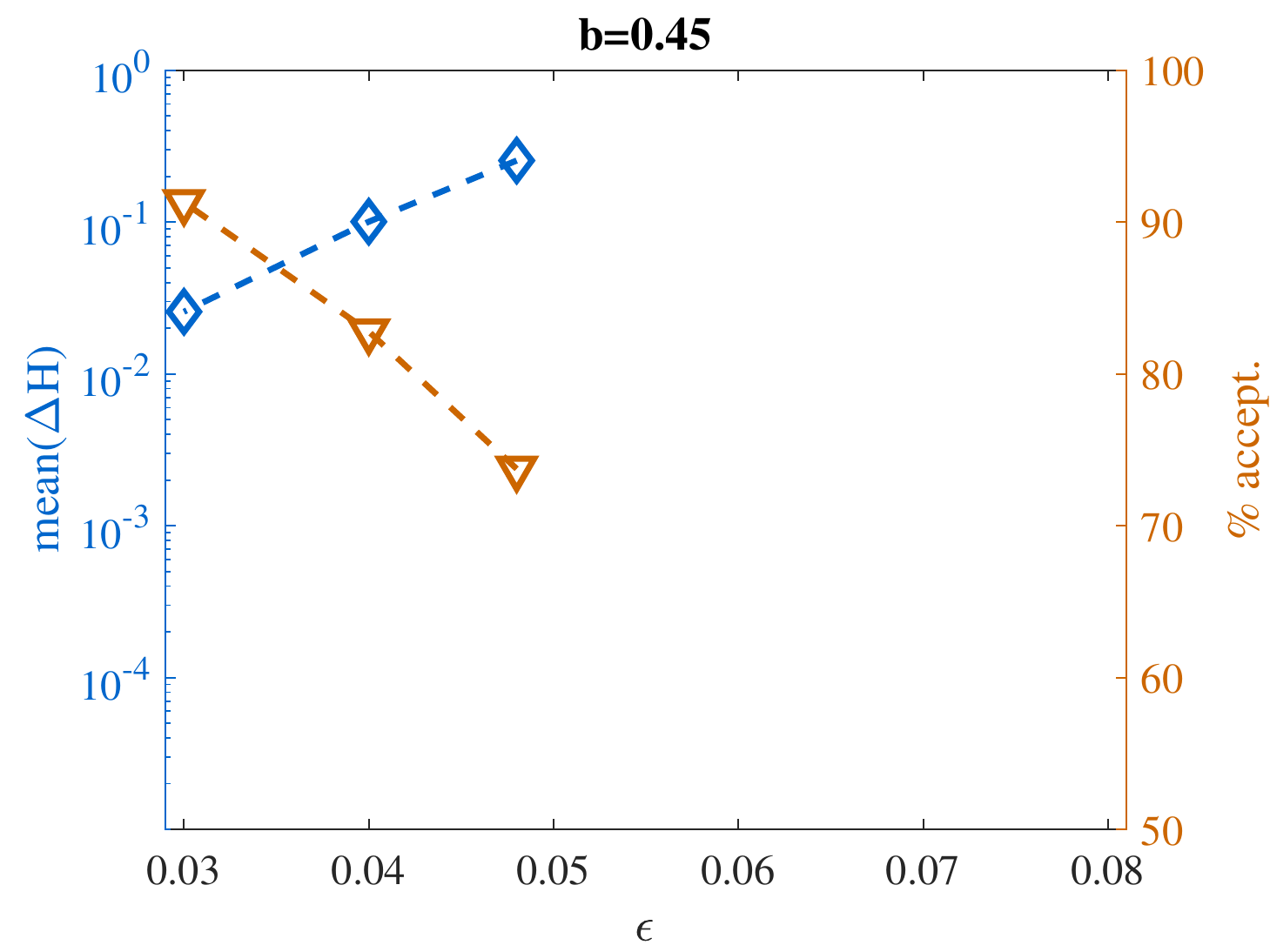}
\caption{Alkane molecule with 15 carbon atoms. Acceptance percentage (triangles) and the  mean of $\Delta H$ (diamonds) as functions of the step-size $\epsilon$, for different values of parameter $b$, when $\tau_{end}=0.48$.}
\label{alkane15}
\end{center}
\end{figure}

Figure~\ref{alkane15}  displays results for \(15\) carbon atoms. We have omitted data corresponding to either acceptance percentages below $45$ or values of the mean of $\Delta H$ larger than 1. It is clear  that LF is far from being the best choice for this problem. BlCaSa and PrEtAl have substantially larger acceptance rates than LF for each value of \(\epsilon\) where LF works. The results for \(5\) and \(9\) carbon atoms are not very different. With \(5\) atoms,  LF, \(b=0.35\), BlCaSa, and PrEtAl can operate with the largest value \(\epsilon = 0.08\), but again, for all values of \(\epsilon\), the acceptance rate of LF is well below that provided by BlCaSa.

As in the preceding test problem, for both BlCaSa and PrEtAl the variation of the mean \(\Delta H\) with \(\epsilon\) is as in the Gaussian model: the former method exhibits a plateau and the second a fast decay (a reduction of more than two orders of magnitude when halving \(\epsilon\) from \(0.06\) to \(0.03\)). This is in spite of the fact that both integrators were derived by considering only the Gaussian case.

\section{Expected acceptance rate vs.\ expected energy error}
\label{sec:mu}
In this section we investigate analytically some properties of the
expected acceptance percentage and the expected energy increments.

\subsection{A formula for the expected acceptance rate}

The following result,  implict in \citep{neal2011}, is needed later.

\begin{lemma}\label{lemma:delta}
Let the integration of the Hamiltonian dynamics \eqref{eq:hamdynamics}
be carried out with a volume-preserving and time-reversible integrator, such that the transformation \(\Psi\) in the phase space that maps the initial condition \((\theta,p)\) into the numerical solution \((\theta^\star,p^\star)\) is continuously differentiable. Assume that \((\theta,p)\sim
\pi(\theta,p)\) (i.e.\ the chain is at stationarity) and that the energy increment \eqref{eq:difference} satisfies
\begin{equation}\label{eq:hypothesis0}
\prob(\Delta H= 0) = 0.
\end{equation}
Then the expected acceptance rate is given by
\[
\E(a) =2\: \prob(\Delta H<0).
\]
\end{lemma}
 \emph{Proof.} Since almost surely \(\Delta H\neq 0\), we may write
\begin{eqnarray*}
  \E(a) &=& \int_{\R^d\times\R^d} \min\big(1, \exp(-\Delta H(\theta,p))\big) \exp(-H(\theta,p))\, d\theta\,dp  \\
   &=& \int_{\Delta H < 0}  \exp(-H(\theta,p))\, d\theta\,dp \\
   && + \int_{\Delta H >0}  \exp(-\Delta H(\theta,p)) \exp(-H(\theta,p))\, d\theta\,dp,
\end{eqnarray*}
and it is enough to show that the last two integrals share a common value.

We denote  by \(\widehat \Psi\) the composition \(S\circ \Psi\), where \(S\) is the momentum flip
\(S(\theta,p) = (\theta,-p)\) (so that \(\widehat \Psi(\theta,p) = (\theta^\star,-p^\star)\)). Then, from the
definition of \(\Delta H\),
\[
\Delta H(\theta,p) +H(\theta,p) = H(\Psi(\theta,p) )
\]
and therefore
\begin{eqnarray*}&&\int_{\Delta H >0}  \exp(-\Delta H(\theta,p)) \exp(-H(\theta,p))\, d\theta\,dp\\
 &&\qquad\qquad\qquad\qquad\qquad= \int_{\Delta H > 0}  \exp(-H(\Psi(\theta,p)))\, d\theta\,dp\\
 && \qquad\qquad\qquad\qquad\qquad= \int_{\Delta H > 0}  \exp(-H(\widehat\Psi(\theta,p)))\, d\theta\,dp,
\end{eqnarray*}
where in the last equality we have taken into account that \(H\circ S=H\). On the other hand,
\begin{eqnarray*}
\int_{\Delta H < 0}  \exp(-H(\theta,p))\, d\theta\,dp & = &
\int_{\Delta H < 0}  \exp(-H(\widetilde\theta,\widetilde p))\, d\widetilde\theta\,\widetilde dp\\
& = & \int_{\Delta H > 0}  \exp(-H(\widehat\Psi(\theta,p)))\, d\theta\,dp.\\
\end{eqnarray*}
 Here the first equality  is just a change in notation.  The second  uses the change of variables \((\widetilde\theta,\widetilde p) = \widehat\Psi(\theta,p)\) which has unit Jacobian determinant because the integrator preserves volume. Pairs \((\widetilde\theta,\widetilde p)\) with \(\Delta H<0\) correspond to pairs \((\theta,p)\) with \(\Delta H>0\) as a consequence of the time-reversibility of the integrator:
 \begin{eqnarray*}
  \Delta H (\widetilde \theta,\widetilde p)  = H(\Psi(\widetilde \theta,\widetilde p)) -H (\widetilde \theta, \widetilde p)= H(S(\theta, p)) - H(S (\Psi(\theta,p)))\qquad&& \\ =H(\theta, p) - H(\Psi(\theta,p))= -\Delta H(\theta,p).\qquad\square&&
 \end{eqnarray*}

Later in the section we shall come across cases where, for simple targets and special choices of \(\epsilon\)
and \(L\), the assumption \eqref{eq:hypothesis0} does not hold. Those cases should be regarded as exceptional
and without practical relevance. In addition,  we point out that the lemma could have been
reformulated without \eqref{eq:hypothesis0}: at stationarity and without counting the proposals with \(\Delta
H(\theta,p)= 0\), one out of two accepted steps comes from proposals with \(\Delta H(\theta,p)< 0\).

\subsection{The standard univariate Gaussian target}

We now investigate in detail the model situation where \(\pi(\theta)\) is the standard univariate
normal distribution and the mass matrix is \(M = 1\), so that \(H = (1/2)p^2+(1/2)\theta^2\). The Hamiltonian system is the standard harmonic oscillator: \(d\theta/d\tau = p\), \(dp/d\tau = -\theta\).
For all time-reversible, volume-preserving, one-step integrators of
practical interest (including implicit Runge-Kutta and splitting integrators), assuming that \(\epsilon\) is in the stability interval, the transformation \((\theta^\star,p^\star) = \Psi(\theta,p)\) associated with an integration leg has an expression \citep{blanes,acta}:
\begin{eqnarray}
\label{eq:method1}\theta^\star &=&\phantom{-\chi^{-1}_\epsilon} \cos(L\alpha_\epsilon) \theta +\chi_\epsilon \sin(L\alpha_\epsilon)p,\\
\label{eq:method2}p^\star  & = &-\chi^{-1}_\epsilon \sin(L\alpha_\epsilon)\theta+\phantom{\chi_\epsilon}\cos(L\alpha_\epsilon) p,
\end{eqnarray}
where \(\chi_\epsilon\) and \(\alpha_\epsilon\) are quantities that depend smoothly on the step-length
\(\epsilon\) and change with the specific integrator, but are independent of the number \(L\) of time-steps.
As \(L\) varies with fixed \(\epsilon\) and \((\theta,p)\), the end point \((\theta^\star,p^\star)\) moves on
an ellipse whose eccentricity is governed by \(\chi_\epsilon\); for a method of order of accuracy \(\nu\),
\(\chi_\epsilon = 1+\mathcal{O}(\epsilon^\nu)\) as \(\epsilon\downarrow 0\). The angle \(\alpha_\epsilon\) is
related to the speed of the rotation of the numerical solution around the origin as \(L\) increases; it
behaves as \(\alpha_\epsilon = \epsilon (1+\mathcal{O}(\epsilon^\nu))\).

From  \eqref{eq:method1}--\eqref{eq:method2}  a short calculation reveals that
 the energy increment has the following expression
\begin{equation}\label{eq:ABC}
2\Delta H (\theta,p) = A \theta^2 + 2 B\theta p + Cp^2,
\end{equation}
with
\begin{eqnarray*}
A &=& \sin^2(L\alpha_\epsilon)(\chi_\epsilon^{-2}-1),\\
B&=& \cos(L\alpha_\epsilon) \sin(L\alpha_\epsilon)(\chi_\epsilon-\chi_\epsilon^{-1}),\\
C&=& \sin^2(L\alpha_\epsilon)(\chi_\epsilon^2-1).
\end{eqnarray*}
For future reference we note that
\begin{equation}\label{eq:discriminant}
B^2-AC = A+C.
\end{equation}

Taking expectations at stationarity in \eqref{eq:ABC},
\begin{equation}\label{eq:AplusC}
2\, \E(\Delta H) = A+C,
\end{equation}
i.e.\
\[
\E(\Delta H) = \sin^2(L\alpha_\epsilon) \rho(\epsilon)
\]
with
\[
\rho(\epsilon) =\frac{1}{2} \left(\chi_\epsilon^2+\chi_\epsilon^{-2}-2\right)= \frac{1}{2} \left(\chi_\epsilon-\frac{1}{\chi_\epsilon}\right)^2\geq 0,
\]
so that, regardless of the choice of \(L\),
\begin{equation}\label{eq:boundwithrho}
0\leq \E(\Delta H) \leq \rho(\epsilon).
\end{equation}
As \(\epsilon\downarrow 0\), \(\rho(\epsilon) = \mathcal{O}(\epsilon^{2\nu})\), and therefore \(\E(\Delta
H)=\mathcal{O}(\epsilon^{2\nu})\), where we remark that the exponent is doubled from the pointwise estimate
\(\Delta H(\theta,p) =  \mathcal{O}(\epsilon^\nu)\), valid for \(\epsilon L =\tau_{end}\). If
\(L\alpha_\epsilon\) happens to be an integer multiple of \(\pi\) then
 \(A= B= C = 0\) and \(\Delta H\equiv 0\); in this case the transformation \(\Psi\) is
  either \((\theta,p)\mapsto (\theta,p)\) or \((\theta,p)\mapsto (-\theta,-p)\) with no energy error
   (but then the Markov chain is not ergodic, this is one of the reasons for randomizing \(\epsilon\)). Also
   note that this gives an example where the hypothesis \(\prob (\Delta H = 0) \neq 0\) in
   Lemma~\ref{lemma:delta} does not hold.

The preceding material has been taken from \cite{blanes}  and we now move to the presentation of new developments.

 For the acceptance rate we have the following result
that shows that \(\E(a)\) is of the form \(\varphi(\E(\Delta H))\) where \(\varphi\) is a monotonically
decreasing  function that does not depend on the integrator, \(\epsilon\) or \(L\).  This result is significant because, even though the
integrator BlCaSa  was derived to minimize  the expected energy error in the univariate standard normal
target, we now show that \emph{it also maximizes the expected acceptance rate}. It is perhaps of some interest to point out that in view of \eqref{eq:acceptance} a reduction of the magnitude of a negative energy error does not change the acceptance probability. 

\begin{theorem}
\label{th:acceptanceunivariate}When the target is the standard univariate Gaussian distribution, the mass matrix is set to \(1\) and the chain is at stationarity, the expected acceptance rate  is given by
\[
\E(a) = 1-\frac{2}{\pi} \arctan \sqrt{\frac{\E(\Delta H)}{2}},
\]
regardless of the (volume-preserving, time-reversible) integrator
 being used, the step-length \(\epsilon\) and the number \(L\) of time-steps  in each integration leg.
\end{theorem}
\emph{Proof.} If in \eqref{eq:ABC}, either  \(\sin(L\alpha_\epsilon)=0\) or \(\chi_\epsilon= 1\), then
\(A=B=C=0\), which leads to \(\Delta H \equiv 0\). In this case \(\E(a) =1\) and \(\E(\Delta H) = 0\) and the
result holds.

If \(\sin(L\alpha_\epsilon)\neq 0\) and \(\chi_\epsilon\neq  1\), then \(A\neq 0\), \(B\neq 0\), \(C\neq 0\)
and \(\Delta H\) vanishes on the straight lines
\[
p = \frac{-B-\sqrt{B^2-AC}}{C} \theta,\qquad p = \frac{-B+\sqrt{B^2-AC}}{C} \theta,
\]
of the \((p,\theta)\) plane with slopes \(m_1\) and \(m_2\) respectively. From the expressions for \(A\) and \(C\) we see that \(AC<0\) and therefore \(m_1m_2=A/C<0\). For the sake of clarity, let us assume that \(C<0\) and \(B>0\), which implies \(m_1>0\) and \(m_2<0\) (the proof in the other cases is similar). Choose angles \(\phi_1\in(0,\pi/2)\), \(\phi_2\in (\pi/2,\pi)\)
with \(\tan(\phi_1) = m_1\), \(\tan(\phi_2) = m_2\).

The energy increment \(\Delta H\) is then negative for points \((\theta,p)\) with polar angles
\(\phi\in(\phi_1,\phi_2)\cup(\phi_1+\pi,\phi_2+\pi)\). Lemma~\ref{lemma:delta} and the symmetry \(H(\theta,p)
= H(-\theta,-p)\) yield
\begin{eqnarray*}
\E(a) &=& 2\int_{\phi\in(\phi_1,\phi_2)\cup(\phi_1+\pi,\phi_2+\pi)}\exp\big(-(1/2)(\theta^2+p^2)\big)\,d\theta\,dp\\
 &=& 4\int_{\phi\in(\phi_1,\phi_2)}\exp\big(-(1/2)(\theta^2+p^2)\big)\,d\theta\,dp.
\end{eqnarray*}
The last integral is easily computed by changing to polar coordinates and then
\[
\E(a) = \frac{2}{\pi} (\phi_2-\phi_1).
\]
We now recall the formula
\[
\arctan(x)-\arctan(y) = \arctan\left(\frac{x-y}{1+xy}\right)
\]
(if the equality sign is understood modulo \(\pi\), then the formula holds for all branches of the multivalued function \(\arctan(x)\)) and, after some algebra find
\[
\E(a) = \frac{2}{\pi} \arctan\left(\frac{2\sqrt{B^2-AC}}{A+C}\right).
\]
Taking into account \eqref{eq:discriminant} and \eqref{eq:AplusC}, the last display may be rewritten in the form
\begin{equation}\label{eq:pole}
\E(a) = \frac{2}{\pi} \arctan\left(\sqrt{\frac{2}{\E(\Delta H)}}\right);
\end{equation}
 the theorem follows from well-known properties of the function \(\arctan(x)\). \(\square\)

In the proof the slopes \(m_1\), \(m_2\) of the lines that separate energy growth from energy decay change
with the integrator, \(\epsilon\) and \(L\); however the angle \(\phi_2-\phi_1\) between the lines only
depends on \(\E(\Delta H)\).

 Note that, as \(\E(\Delta H)\downarrow 0\),
\begin{equation}\label{eq:origin}
\E(a) = 1- \frac{\sqrt{2}}{\pi} \sqrt{\E(\Delta H)}+\mathcal{O}\big((\E(\Delta H))^{3/2}\big),
\end{equation}
and, as a consequence, as \(\epsilon\downarrow 0\), \(L\uparrow \infty\) with \(L\epsilon=\tau_{end}\),
\[
\E(a) = 1-\mathcal{O}(\epsilon^\nu)
\]
for an integrator of order of accuracy \(\nu\). On the other hand, the formula \eqref{eq:pole} shows that, as
\(\E(\Delta H) \uparrow \infty\), the expected acceptance rate decays slowly, according to the estimate,
\begin{equation}\label{eq:decay}
\E(a) \sim \frac{2\sqrt{2}}{\sqrt{\E(\Delta H)}}.
\end{equation}
For instance, if \(\E(\Delta H) = 100\), then \eqref{eq:pole} yields \(\E(a)\approx 0.089\); this should
perhaps be compared with the data in the right panel of Figure~\ref{Gaussian_fig4_256_1024}, where the target is low dimensional.

We conclude our analysis of the standard normal by presenting a result that will be required later. It is
remarkable that the formulas in this lemma express moments of \(\Delta H\) as polynomials in \(\E(\Delta H)\)
and that, furthermore, those polynomials do not change with the value of \(\epsilon\), \(L\) or with the
choice of integrator.

\begin{lemma}\label{le:moments} For a univariate, standard Gaussian target with
unit mass matrix and a time-reversible, volume preserving integrator, setting \(\mu = \E(\Delta H)\), we have:
\begin{eqnarray*}
\E((\Delta H)^2) &=& 2\,\mu+3\,\mu^2,\\
\E((\Delta H)^3) &=& 18\,\mu^2+15\,\mu^3, \\
\E((\Delta H)^4) &=& 36\, \mu^2+180\,\mu^3+105\,\mu^4.
\end{eqnarray*}
\end{lemma}
\emph{Proof.} Raise \eqref{eq:ABC} to the second (third or fourth) power. Compute expectations to write
\(\E((\Delta H)^2)\) (\(\E((\Delta H)^3)\) or \(\E((\Delta H)^4)\))  as a polynomial in \(A, B, C\).  Eliminate odd powers of \(B\) by integration of the Gaussian distribution. Then
express
\(B^2\) in terms of \(A\) and \(C\) by using \eqref{eq:discriminant} and a little  patience. Finally use
\eqref{eq:AplusC}. \(\square\)

As \(\mu\rightarrow 0\), \(\E((\Delta H)^2) / \E(\Delta H)\rightarrow 2\), in agreement with Proposition 3.4
in \cite{optimal}. In addition, also \(\E((\Delta H)^4) / \E((\Delta H)^3)\rightarrow 2\), a result that,
while not contained in the paper by \cite{optimal} may be proved with the tools used to prove that
proposition.

\subsection{Univariate Gaussian targets}

For the univariate problem where \(\pi(\theta) = {\cal N}(0,\sigma^2)\) and \(M=1\), the Hamiltonian is
\((1/2)p^2+(1/2)(\theta^2/\sigma^2)\) and the differential equations are \(d\theta/d\tau = p\), \(dp/d\tau =
-\theta/\sigma^2\). This problem may be reduced to the case \(\pi(\theta) = {\cal N}(0,1)\), \(M=1\) by
scaling the variables, because for all integrators of practical interest the operations of rescaling and
numerical integration commute. Specifically, the new variables are \(\theta/\sigma\), \(\tau/\sigma\) with
\(p\) remaining as it was.

By changing variables we find that
the bound \eqref{eq:boundwithrho} should be replaced by
\begin{equation}\label{eq:boundwithrhobis}
0\leq \E(\Delta H) \leq \rho(\epsilon/\sigma).
\end{equation}
(The bound \eqref{eq:bound} is a direct consequence of this.) On the other hand, for arbitrary \(\sigma\),
Theorem~\ref{th:acceptanceunivariate} and
Lemma~\ref{le:moments} hold as they stand.

\subsection{Multivariate Gaussian targets}
We now consider for each \(d=1,2,\dots\) a centered Gaussian target
 \(\pi_d(\theta)\), \(\theta\in\R^d\) with  covariance matrix \(C_d\). The  \(C_d\) are assumed
 to be nonsingular, but otherwise they are allowed to be completely arbitrary. As pointed out before we may
 assume that the \(C_d\) have been diagonalized.
We run HMC  for each \(\pi_d(\theta)\) and, for simplicity, assume that the unit mass matrix \(I_d\) is used
(but the result may be adapted to an arbitrary  constant mass matrix). Then \(\pi_d(\theta)\) is a product of univariate
distributions \(\pi_{d,\ell}(\theta_\ell)\), \(\ell=1,\dots,d\),  the Hamiltonian \(H_d\) is a sum of
one-degree-of-freedom Hamiltonians \(H_{d,\ell}\) and the increment \(\Delta H_d(\theta,p)\) is a sum of
increments \(\Delta H_{d,\ell}(\theta_\ell,p_\ell)\). The integrator being used for \(\pi_d(\theta)\) is
allowed to change with \(d\); it is only assumed that when applied to the standard normal univariate target
takes the form \eqref{eq:method1}--\eqref{eq:method2}. Similarly the step-length \(\epsilon\) and the length
\(\tau_{end}\) of the integration interval are allowed to change with \(d\).

Our next result shows that, even in this extremely general scenario,
simple hypotheses on the expectations of
the
 \(\Delta H_{d,\ell}\) ensure that, asymptotically,
 the variables \(\Delta H_d\) are  normal with a variance that is twice as
 large as the mean \citep{neal2011}.\footnote{This relation between the variance and the mean follows heuristically from the following argument, that has appeared many times in the literature. As in Lemma~\ref{lemma:delta}, it is easily proved that \(\E(\exp(-\Delta H))=1\). Then \(\E(\exp(-\Delta H)) \simeq 1-\E(\Delta H)+(1/2) \E((\Delta H)^2).\)}

\begin{theorem}\label{th:central}In the general scenario  described above, assume that as \(d\uparrow \infty\)
\begin{equation}\label{eq:hypothesis1}
M_d := \max_{1\leq \ell \leq d} \E(\Delta H_{d,\ell}) \rightarrow 0
\end{equation}
and, for some \(\mu\in [0,\infty)\),   as \(d\uparrow \infty\)
\begin{equation}\label{eq:hypothesis2}
\E(\Delta H_d) = \sum_{\ell = 1}^d \E(\Delta H_{d,\ell}) \rightarrow \mu.
\end{equation}

Then, as \(d\uparrow \infty\):
 \begin{itemize}\item The  distributions of the random variables \(\Delta H_d\) at stationarity converge to the distribution \({\cal N}(\mu,2\mu)\).

\item The acceptance rates \(a_d\) for the targets  \(\pi_d(\theta)\), \(d=1,2,\dots\), satisfy
\[
\E(a_d) \rightarrow 2 \Phi(-\sqrt{\mu/2}).
\]
\end{itemize}
\end{theorem}
\emph{Proof.} The first item is a direct application of the central limit theorem. From
Lemma~\ref{le:moments}, the variance of \(\Delta H_d\) is, with \(\mu_{d,\ell}:= \E(\Delta H_{d,\ell})\),
\[
s^2_d = \sum_{\ell=1}^d (2 \mu_{d,\ell}+2 \mu_{d,\ell}^2),
\]
which tends to \(2\mu\) because, for any integer \(k>1\),
\begin{equation}\label{eq:auxiliar}
\sum_{\ell=1}^d  \mu_{d,\ell}^k \leq M_d^{k-1}\: \sum_{\ell=1}^d  \mu_{d,\ell}.
\end{equation}

For the validity of the central limit theorem, instead of the more common Lindeberg condition,  we will
 check the well-known  Lyapunov condition based on controlling the centered fourth moment. We then consider the expression
\[
\frac{1}{s_d^4}\sum_{\ell=1}^d \E\big((\Delta H_{d,\ell}-\mu_{d,\ell})^4\big).
\]
We expand the fourth power and use Lemma~\ref{le:moments} to write
 the expectation as a linear combination of \(\mu_{d,\ell}^2\), \(\mu_{d,\ell}^3\), \(\mu_{d,\ell}^4\)
 with coefficients that are independent of \(d\) and \(\ell\). Then  the sum in the display tends to 0  in view of \eqref{eq:auxiliar} with \(k=2,3,4\), and accordingly the Lyapunov condition holds.

For the second item,  since    \(u\mapsto \min(1,\exp(-u))\) is bounded and continuous, we may write
\[
\E(a_d) \rightarrow \E\big(\min(1,\exp(-\zeta)\big),
\]
where \(\zeta\sim {\cal N}(\mu,2\mu)\). The proof concludes by applying Lemma~\ref{lemma:three} below.
\(\square\)
\smallskip

The following result, whose proof is an elementary computation and will be omitted,  has been
 invoked in the proof of Theorem~\ref{th:central}. The first equality should be compared with
 Lemma~\ref{lemma:delta}.

\begin{lemma}\label{lemma:three}
If \(\zeta \sim {\cal N}(\mu,2\mu)\), then
\[
\E\Big(\min\big(1,\exp(-\zeta)\big)\Big) = 2\prob(\zeta<0) = 2\Phi(-\sqrt{\mu/2}).
\]
\end{lemma}

As an application of Theorem~\ref{th:central}, we review our findings in
Figure~\ref{Gaussian_fig4_256_1024}.\footnote{The experiments for that figure had randomized step-lengths.
Even though the effect of randomization may be easily analyzed, we prefer to ignore this issue in order to
simplify the exposition.} In the scenario above, \eqref{eq:gaussianexample} plays the role of
\(\pi_d(\theta)\).  We set
 \(\tau_{end} = 5\) and for each \(d\) choose a step-length \(\epsilon_d\) and one of the six integrators.
 According to \eqref{eq:boundwithrhobis}, \(\E(\Delta H_{d,\ell})\leq C \ell^4\epsilon_d^4\), with \(C\) the largest of the constants corresponding to the different integrators tested.
  Assumption \eqref{eq:hypothesis1} will hold if \(\epsilon_d = o(d^{-1})\). Furthermore, if \(\epsilon_d\) decays like
\(d^{-5/4}\), the sum in the left-hand side of \eqref{eq:hypothesis2} will be bounded above by a constant
multiple of
\[
\sum_{\ell=1}^d \ell^4 \epsilon_d^4 =  \mathcal{O}(1),
\]
and one may prove that \eqref{eq:hypothesis2} holds for some \(\mu<\infty\). Therefore, for \(d\) large,
\(\Delta H_d(\theta,p)\) will be asymptotically \({\cal N}(\mu,2\mu)\). In fact the  curve in
Figure~\ref{Gaussian_fig4_256_1024}  is given by the equation
 \(a= 2\Phi(-\sqrt{\mu/2})\). Of course if, as \(d\) varies, \(\epsilon\) decays faster than \(d^{-5/4}\)
 then the expectation of the acceptance rate would approach \(100\%\); a slower decay would imply that the
 expectation would approach \(0\).

The equation \(a=2\Phi(-\sqrt{\mu/2})\) is well known \citep{neal2011} and has been established rigorously by
\cite{optimal} in a restrictive scenario. There the target is assumed to be a product of \(d\) independent
copies of \emph{the same}, possibly multivariate, distribution \(\pi^\star\). The integrator and
\(\tau_{end}\) are not allowed to vary with \(d\) and \(\epsilon_d\) has to decrease as \(d^{-1/4}\). Some of
these assumptions are far more restrictive than those we required above. On the other hand, \(\pi^\star\) is
\emph{not} assumed to be Gaussian. Thus there are relevant differences between the analysis by \cite{optimal}
and the present study. However the underlying idea is the same in both treatments: one considers the energy
error as a sum of many independent small contributions and for each contribution the variance is,
approximately, twice as large as the mean. The role played by Proposition 3.4 in the paper by \cite{optimal}
is played here by  Lemma~\ref{le:moments}, where we saw that  the second moment \(\E\big((\Delta H)^2\big)\)
(and by implication the variance) is approximately twice as large as the expectation \(\E(\Delta H)\) when the
latter is small.

The function \(2 \Phi(-\sqrt{\mu/2})\) decays exponentially as \(\mu\uparrow \infty\) (to be compared with
\eqref{eq:decay}) and, as \(\mu\downarrow 0\) behaves as
\[
1-\frac{1}{2\sqrt{\pi}}\sqrt{\mu}+\mathcal{O}(\mu^{3/2})
\]
(to be compared with \eqref{eq:origin}).

\subsection{Other targets}
\begin{figure}
\includegraphics[width=0.48\hsize]{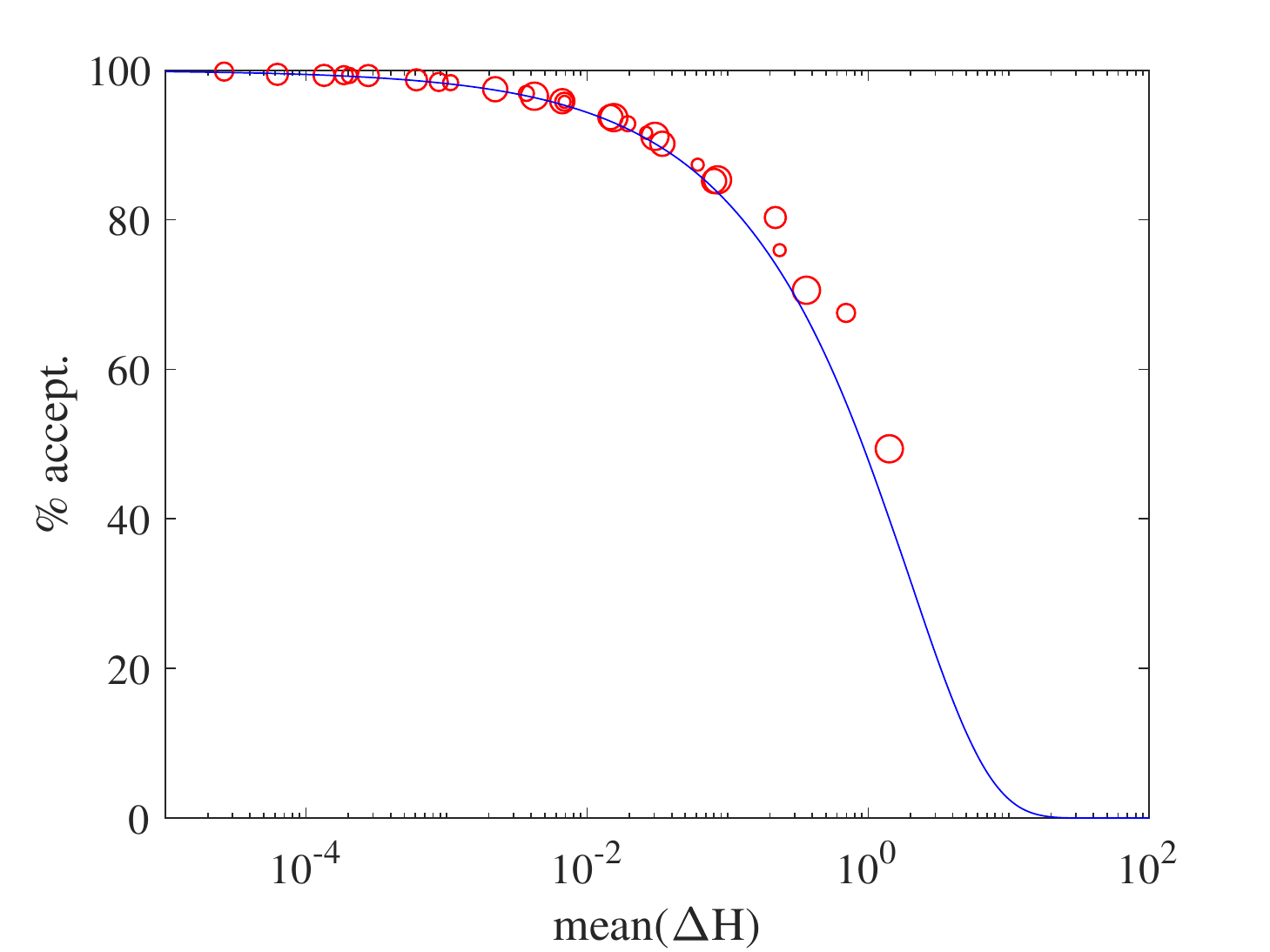} \quad
\includegraphics[width=0.48\hsize]{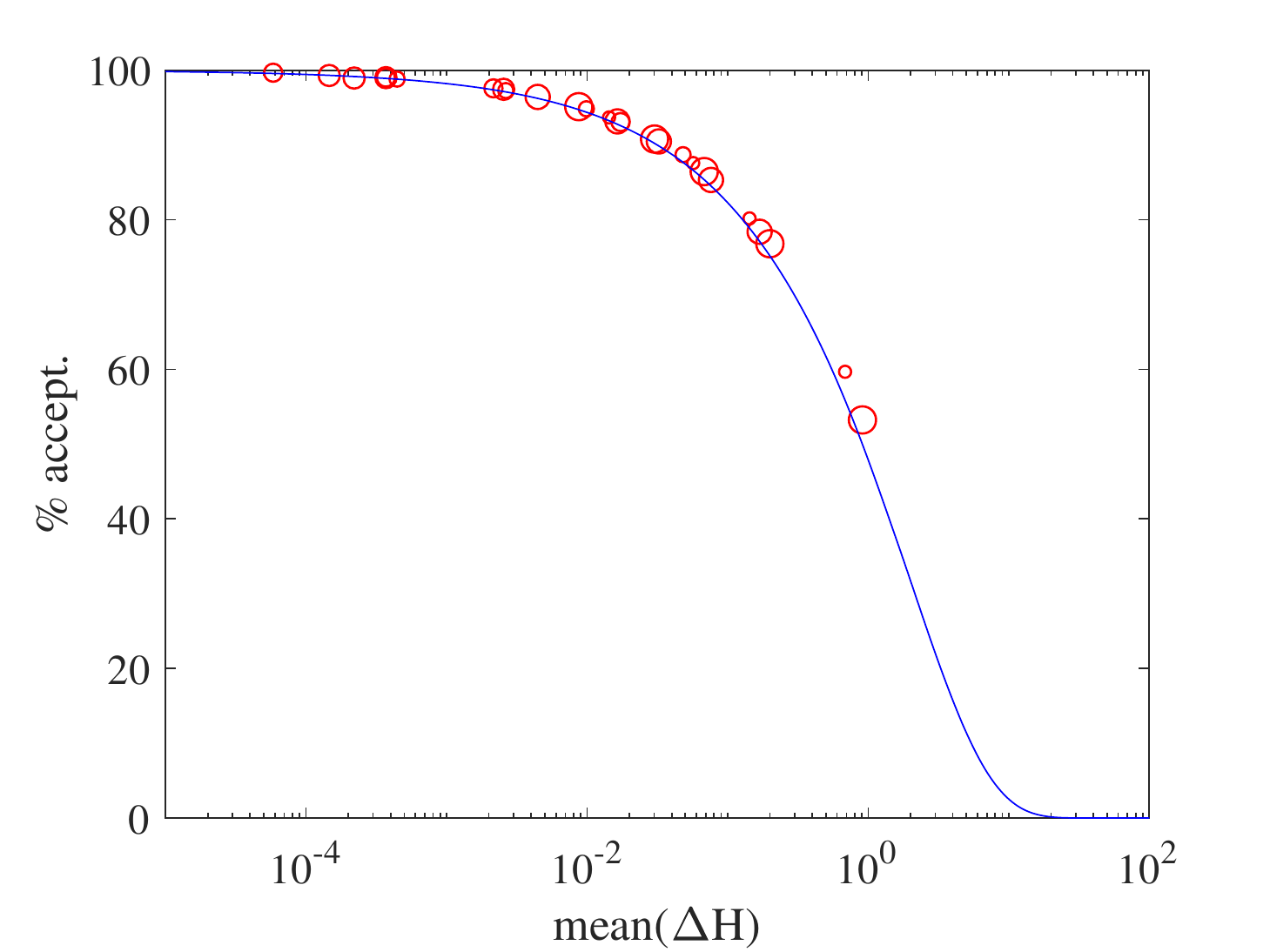}
\\
\includegraphics[width=0.48\hsize]{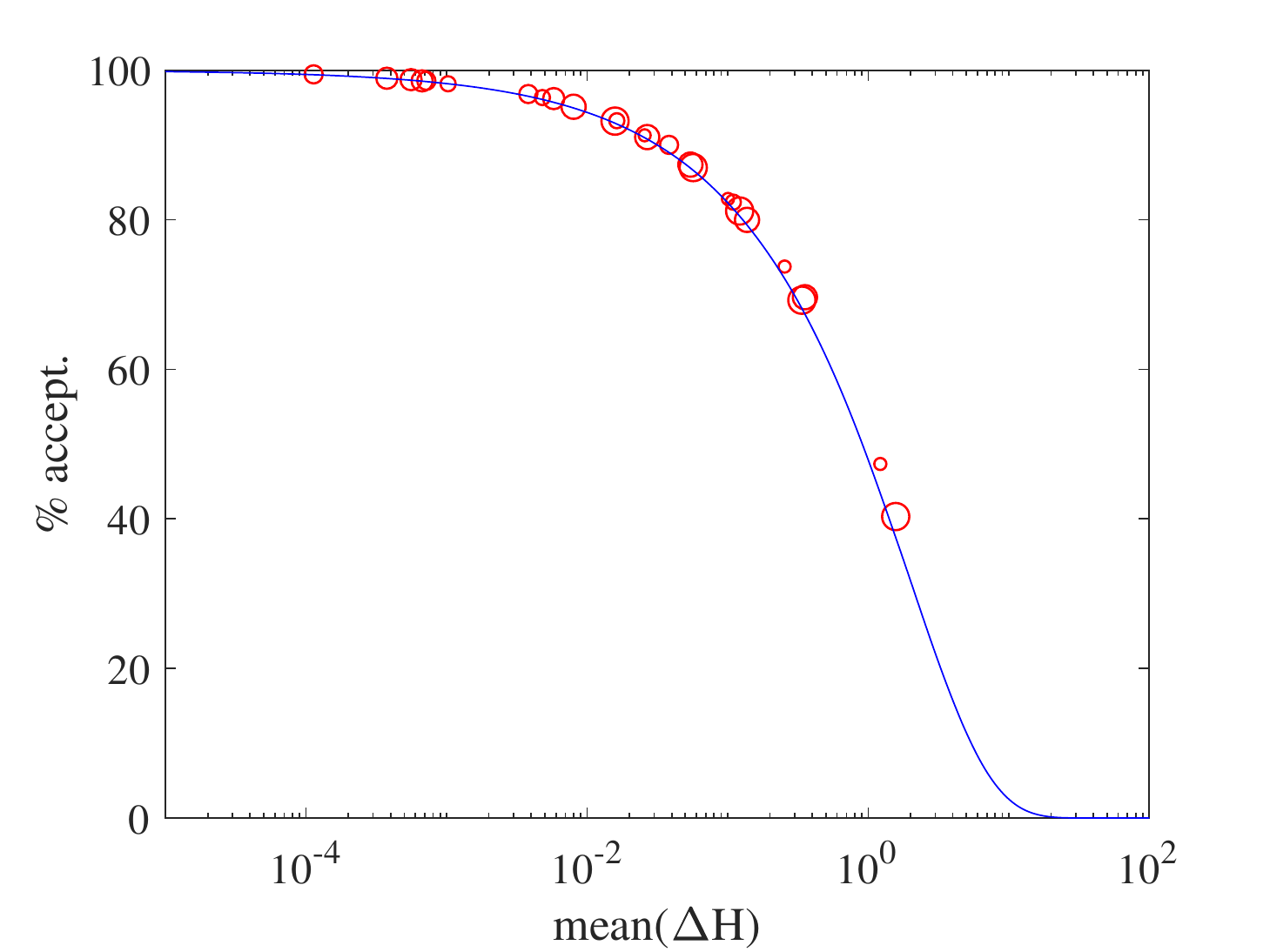} \quad
\includegraphics[width=0.48\hsize]{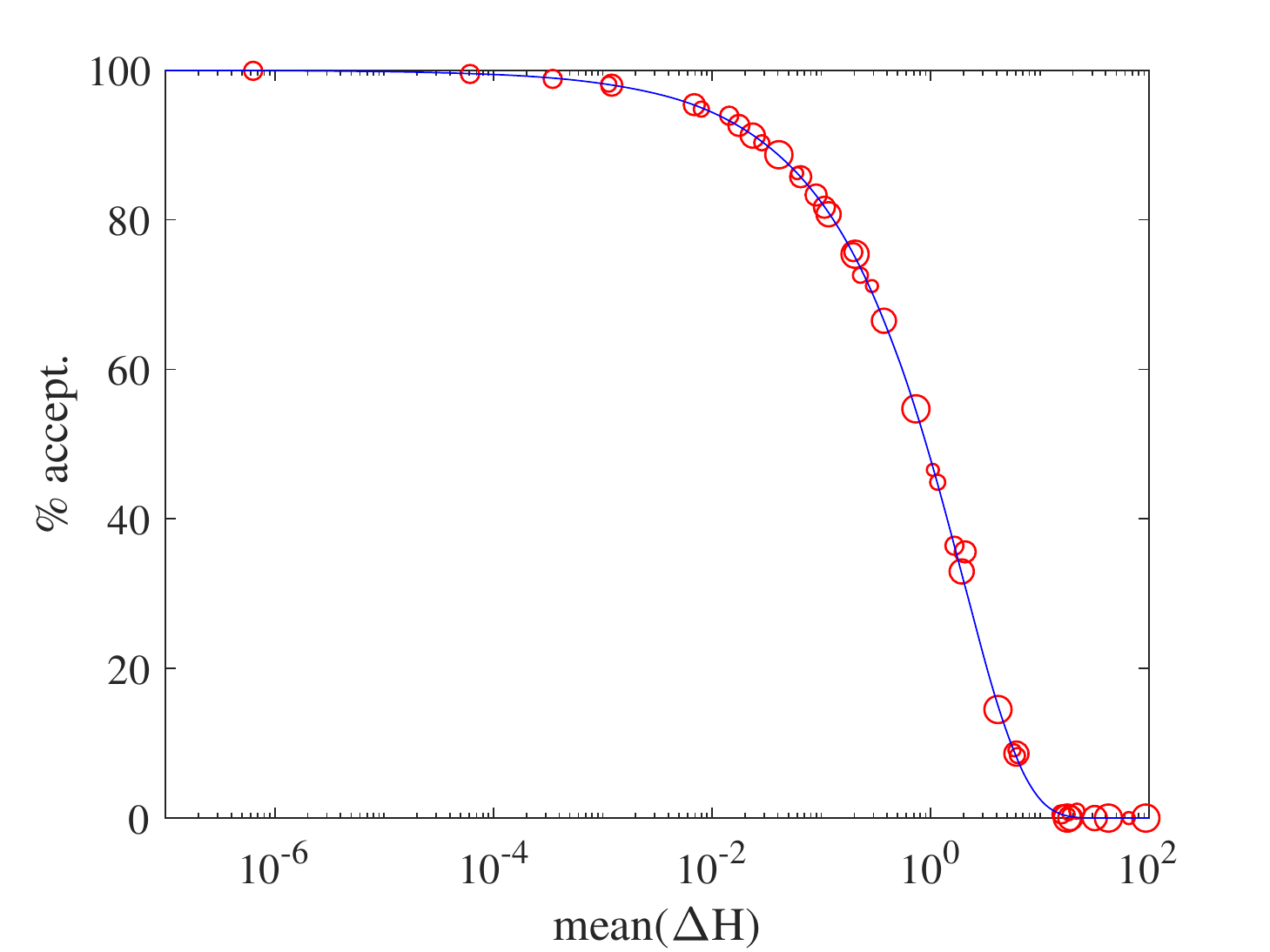}
\caption{Alkane molecule with 5 carbon atoms (top left),  9 carbon atoms (top right) and 15 carbon atoms (bottom left), Log Gaussian Cox problem (bottom right). The dimensions \(d\) are 15, 27, 45 and 4096, respectively. Acceptance percentage vs.\  energy error. Points come from all six integrators and all step-lengths. The diameter of the circles indicates the integrator used.  The continuous line corresponds to the case where \(\Delta H\) is \({\cal N}(\mu,2\mu)\).}
\label{Cox_alkane_fig4}
\end{figure}

Our final Figure~\ref{Cox_alkane_fig4} collects information from all six integrators
and all step-lengths for the molecules with 5, 9 or 15 carbon atoms and the Log-Gaussian Cox problem.
 The points  almost lie
on the curve  \(a=2\Phi(-\sqrt{\mu/2})\). Moreover, the fit gets better as the dimensionality increases.
 In other words, we experimentally observe here a behaviour that reproduces our theoretical analysis
  of the \emph{Gaussian} case. This is in line with our earlier findings about the change of
   the average energy error as \(\epsilon\) varies. Thus the experiments in this paper seem
    to show that, for \(d\) large, properties of HMC for Gaussian targets also hold
    for general targets (or at least for a large class of relevant non-Gaussian targets).  A possible explanation for this behaviour is that, in the tests reported, the algorithm expends much time in the neighbourhood of one mode, where the target may be approximated by a Gaussian.

\section{Conclusions}
\label{sec:conclusions} Numerical experiments have proved that, for HMC sampling from probability
distributions in \(\R^d\), \(d\gg 1\), the integrator suggested by \cite{blanes}, while being as easy to
implement as standard leapfrog, may deliver three times as many accepted samples as leapfrog with the same computational cost.

In addition we have given some theoretical results in connection with the acceptance rate of HMC at
stationarity. We proved rigorously  a central limit theorem behavior in a very general scenario of Gaussian
distributions. We also provided a detailed analysis of the model problem given by the univariate normal
distribution; this analysis implies that the  integrator in \cite{blanes}, while derived to minimize the
expected energy error, also maximizes the expected acceptance rate.
\bigskip

{\bf Acknowledgements.}  M.P.C. has been supported by projects PID2019-104927GB-C22 (GNI-QUAMC), (AEI/FEDER, UE)
 VA105\-G18 and VA\-169P20 (Junta de Castilla y Leon, ES) co-financed by
FEDER funds. J.M.S.-S. has  been supported by project
  PID2019-104927GB-C21 (AEI/FEDER, UE).
D.S.-A. was supported by the NSF Grant DMS-2027056 and the NSF Grant DMS-1912818/ 1912802. J.M.S.  would like to thank the Isaac Newton
Institute for Mathematical Sciences (Cambridge) for support and hospitality during the programme “Geometry,
compatibility and structure preservation in computational differential equations” when his work on this paper
was completed.

\bibliography{bibliographyMP}
\bibliographystyle{plainnat}
\end{document}